\documentclass[11pt]{article}

\usepackage[T1]{fontenc}
\usepackage[utf8]{inputenc}
\usepackage{lmodern}
\usepackage{amsmath}
\usepackage{amsfonts}
\usepackage{amssymb}
\usepackage{amsthm}
\usepackage{enumitem}
\usepackage{geometry}
\usepackage{microtype}
\usepackage{authblk}
\usepackage[round,sort]{natbib}
\usepackage{xcolor}
\usepackage[colorlinks=true,citecolor=blue!55!black,linkcolor=blue!55!black,urlcolor=blue!55!black]{hyperref}

\usepackage{mathpazo}

\newtheorem{theorem}{Theorem}
\newtheorem{lemma}{Lemma}
\newtheorem{proposition}{Proposition}
\newtheorem{corollary}{Corollary}
\theoremstyle{definition}
\newtheorem{definition}{Definition}
\newtheorem{example}{Example}
\newtheorem{remark}{Remark}

\newcommand{\alt}{\mathcal{A}}
\newcommand{\linalt}{\mathcal{P}}

\newcommand{\Sc}{\mathrm{Sc}}
\newcommand{\pri}{\mathrel{\triangleright}}
\newcommand{\bt}{\mathbf{t}}
\newcommand{\br}{\mathbf{r}}

\title{\textbf{Obvious group manipulations in voting}\thanks{An earlier version of this
project circulated under the title ``Obvious manipulations by groups.'' We are grateful to the participants of the Social Choice and Welfare online seminar, the MLSE Seminar at Maastricht University, and the COMSOC Seminar at the University of Amsterdam for comments and suggestions that led to improvements of the paper. Federico Fioravanti acknowledges financial support from the French National Research Agency within the project ANR-24-EXMA-0001 PEPR MathsVivES CONDORCET.}}
\author[1]{R. Pablo Arribillaga}
\author[1,2]{Agustín G. Bonifacio}
\author[2]{Federico Fioravanti\footnote{federico.fioravanti@univ-st-etienne.fr (corresponding author).}}
\affil[1]{Instituto de Matem\'{a}tica Aplicada San Luis, UNSL and CONICET,
San Luis, Argentina.}
\affil[2]{Universit\'e Jean Monnet Saint-\'Etienne, CNRS, Universit\'e Lyon 2, emlyon business school, GATE, 42023, Saint-\'Etienne, France.}

\begin{document}

\maketitle

\begin{abstract}
We study obvious group manipulation in voting and characterize it
as manipulation identifiable from a coalition's option sets alone.
Our analysis focuses on two overlapping classes: tops-only rules
and rules satisfying the majority criterion, including all
Condorcet-consistent rules. On the universal domain with at least
three alternatives, a tops-only rule is not obviously manipulable
by groups (NOMG) if and only if it is dictatorial. For rules
satisfying the majority criterion, NOMG is impossible when
$n,m\geq3$, except with four voters and three alternatives.
We quantify the coalition sizes behind these restrictions,
obtaining sharp bounds for efficient tops-only rules and an exact
threshold for majority-criterion rules with odd electorates.
In the exceptional configuration, we characterize an interval
of NOMG scoring rules containing Borda, which is also Condorcet
consistent there. We further characterize NOMG with two
alternatives and apply our tops-only characterization to
generalized median voter schemes and voting by committees.
\end{abstract}



\section{Introduction}\label{sec:intro}

The Gibbard--Satterthwaite theorem
\citep{gibbard1973manipulation,satterthwaite1975strategy} makes manipulation
unavoidable for every onto and non-dictatorial voting rule on the universal
domain when there are at least three alternatives. It does not distinguish
between a deviation that is easy to recognize and one whose success depends
on fine information about the other voters. The notion of obvious
manipulation, introduced by \citet{TROYAN2020104970}, makes this distinction
by comparing the best and the worst outcomes that a report can generate. It
has been used to discriminate among manipulable voting rules
\citep{azizobviousmanipulations2021,ARRIBILLAGA202412}: many familiar rules
are manipulable, but not obviously so.

This paper asks what changes when manipulation is coordinated by a
coalition. Coalitions matter because many rules that pass the individual
test owe their good behavior to the fact that a single voter has little
influence. The information a coalition needs in order to recognize a
profitable deviation is of the same kind as for an individual: it must know
which outcomes its joint report can produce, not how the outsiders will
vote.

Given a coalition, its members' true preferences, and a misreport, we
compare the set of outcomes left open by truthful reporting with the set
left open by the misreport. The misreport is an \emph{obvious group
manipulation} if either some outcome left open by truthful reporting is
ranked by every member below every outcome left open by the misreport, or
some outcome left open by the misreport is ranked by every member above
every outcome left open by truthful reporting. For a single voter this is
the notion of \citet{TROYAN2020104970}, and for the grand coalition it is an
ordinary group manipulation. A rule is \emph{not obviously manipulable by
groups} (NOMG) if no coalition has an obvious group manipulation.

We first give an informational foundation for this definition: with at least
three alternatives, obvious group manipulations are exactly the misreports that
remain group manipulations under every voting rule generating the same option
sets for the coalition (Proposition~\ref{prop:indistinguishability}). This is the coalitional
counterpart of the characterization in \citet{TROYAN2020104970};
\citet{doi2026obvious} independently introduced the same notion and
obtained this result in a general mechanism-design framework. The result explains why the definition captures
deviations whose profitability can be recognized from option sets alone:
for each rule consistent with this information, the deviation benefits
every coalition member at some outsider profile, although that profile
may differ across rules. Conversely, whenever a deviation is not obvious,
there is a rule with the same option sets under which it never benefits
all coalition members. Thus, the definition has a precise informational
interpretation beyond its formal extension of the individual notion.

Within the universe of voting rules, we focus on two prominent
classes that capture different aspects of collective choice.
Tops-only rules restrict the preference information used by the
rule: voters' favorite alternatives suffice to determine the
outcome. This class includes plurality, generalized median voter
schemes, and voting by committees. Its importance also stems from
the connection between informational simplicity and incentives:
\citet{chatterji2011tops} study domains on which strategy-proofness
and unanimity imply the tops-only property. Rules satisfying the
majority criterion instead restrict the response to concentrated
support: an alternative ranked first by a strict majority must
be selected, while lower ranks may otherwise affect the outcome.
This class includes every Condorcet-consistent rule. The broader
axiomatic case for majority-based aggregation is developed by
\citet{dasguptaMaskin2008,dasguptaMaskin2020}, who study the
robustness and characterization of majority rule across
preference domains.

These classes provide complementary perspectives on coalitional
manipulation: one concerns which preference information a rule
uses, and the other concerns the power of a strict majority.
They overlap---plurality belongs to both---and neither contains
the other. Their comparison therefore lets us examine how
obvious group manipulation arises under two distinct restrictions,
and how each restriction shapes the protection available against
coalitions of different sizes.

Our remaining contributions establish both the restrictions imposed by
coalitional obviousness and the possibilities for protection against it.
With two alternatives, we characterize NOMG and show that it can be
strictly weaker than strategy-proofness. With at least three alternatives,
we characterize immunity for tops-only rules and show that full immunity
within this class implies dictatorship. Beyond tops-only rules, we
establish an impossibility for rules satisfying the majority criterion.
We then compare the coalition sizes behind these restrictions, obtaining
sharp bounds for efficient tops-only rules and an exact threshold for
rules satisfying the majority criterion with odd electorates. An exceptional family of scoring
rules provides a positive result, for which we characterize NOMG exactly. Finally, we apply
the tops-only characterization to two familiar families of rules,
identifying the parameters that provide the greatest protection
against obvious group manipulation.

We first consider the case of two alternatives. The only coalitions
that can gain unanimously are those whose members agree on their
preferred alternative. An efficient rule is NOMG precisely when
every proper coalition that can force an alternative can do so
by voting unanimously for it
(Proposition~\ref{prop:binary}). This condition is much weaker
than strategy-proofness: parity rules satisfy it. Thus, with
two alternatives, immunity to obvious group manipulation
allows for a broader class of rules than strategy-proofness.

With at least three alternatives, we begin with tops-only rules.
For every coalition, we characterize when a tops-only rule leaves it
without an obvious group manipulation
(Theorem~\ref{thm:characterization}). For efficient rules the
characterization is a single condition, extending the individual
veto condition of \citet{ARRIBILLAGA202412} to coalitions:
if a coalition can veto an alternative with some report, it
vetoes it with every report at which none of its members ranks
that alternative first.

A tops-only rule is NOMG if and only if it is group strategy-proof
(Theorem~\ref{thm:gsp}). The Gibbard--Satterthwaite theorem then implies
that only dictatorships qualify. For tops-only rules, therefore, the weak
informational requirement of obviousness has exactly the same bite as full
group strategy-proofness. The proof shows that the supporters of a winning
alternative always include a coalition that can force it; a coalition whose
members all prefer a different outcome is disjoint from it and therefore
leaves the winner in place.

We then extend the analysis beyond tops-only rules.
Every rule satisfying the majority criterion, and hence every
Condorcet-consistent rule, is obviously manipulable by groups whenever
$n,m\geq3$ and $(n,m)\neq(4,3)$
(Theorem~\ref{thm:majority-omg}). Thus, allowing a rule to use information
beyond voters' top alternatives does not restore compatibility between
NOMG and the majority criterion, except possibly in a single configuration
of electorate and alternative-set sizes.

We next determine how large coalitions must be for these restrictions
to become binding. We say that a rule is \emph{$k$-NOMG} if no coalition
with at most $k$ members has an obvious group manipulation. The grand
coalition contributes
only efficiency: a rule is NOMG if and only if it is efficient and no
proper coalition has an obvious group manipulation. For efficient
tops-only rules, immunity to coalitions of up to $n-1$ voters implies
dictatorship (Proposition~\ref{prop:sizes}), and this bound is sharp: for every $n\geq3$, the rule at
which a designated voter decides unless all other voters agree on a
different alternative is efficient, non-dictatorial, and immune to
coalitions of up to $n-2$ voters. Thus, the impossibility of full immunity
is compatible with protection against all but the largest coalitions.
Without efficiency, even immunity to all proper coalitions is compatible
with non-dictatorial rules.

For rules satisfying the majority criterion, and hence for
Condorcet-consistent rules, every coalition smaller than half the
electorate is protected, whereas some minimal strict majority has
an obvious group manipulation whenever $n,m\geq3$ and
$(n,m)\neq(4,3)$ (Proposition~\ref{prop:majority-sizes}). Thus, for
odd $n$, the smallest manipulating coalition has exactly $(n+1)/2$
members; for even $n$, its size is either $n/2$ or $n/2+1$.
For plurality, which is both tops-only and satisfies the majority
criterion, the smallest such coalition has $\lceil n/2\rceil$
members (Proposition~\ref{prop:plurality}).

The configuration excluded from the majority results yields a positive result. When $(n,m)=(4,3)$, the scoring
rule with vector $(1,t,0)$ is NOMG if and only if
$\tfrac12\leq t<\tfrac23$ (Theorem~\ref{thm:n4m3}). This interval contains
Borda's rule and consists of rules satisfying the majority criterion.
Borda's rule is even Condorcet consistent in this electorate
(Corollary~\ref{cor:borda}). These rules show that full immunity to obvious
group manipulation can coexist with anonymity, efficiency, and majority
responsiveness on the universal domain. They also establish that the
tops-only restriction is essential for our dictatorship result and that
the exception in the majority impossibility theorem is substantive.

Finally, we apply the tops-only characterization to two classes of rules
that are strategy-proof on restricted domains: generalized median
voter schemes, strategy-proof on single-peaked preferences
\citep{moulin1980strategy}, and voting by committees,
strategy-proof on separable preferences \citep{barbera1991voting}.
On the universal domain, only their dictatorial members are NOMG, so the
question is against which coalition sizes the others are protected.
A generalized median voter scheme is $k$-NOMG if and only if two conditions
on its fixed ballots hold for every coalition with at most $k$ members:
such a coalition can veto at most the two extreme alternatives, and it
cannot exclude both of them by splitting its reports between them
(Theorem~\ref{thm:gmvs}). Among non-dictatorial generalized median voter
schemes, the largest coalition size that can be protected is the largest
integer strictly below $n/2$ when $m\geq4$, and strictly below $2n/3$ when
$m=3$; anonymous median voter schemes attain both bounds
(Proposition~\ref{prop:median}). A voting by committees is $k$-NOMG if and
only if no coalition with at most $k$ members can force an object into or
out of the outcome (Theorem~\ref{thm:committees}), which requires $k<n/2$;
among voting by quota, simple majority on each object gives the greatest
protection (Corollary~\ref{cor:quota}). For $k=1$, these characterizations
recover the corresponding results for individual obvious manipulation.

\paragraph{Related literature.}  Obvious manipulations were introduced by
\citet{TROYAN2020104970}, who show that a manipulation is obvious exactly
when it can be recognized as profitable by a cognitively limited agent in the
sense of \citet{li2017obviously}, and who use the notion to classify
non-strategy-proof mechanisms in school choice, auctions, and bilateral trade.
In voting, \citet{azizobviousmanipulations2021} identify classes of rules that are obviously manipulable, and
\citet{ARRIBILLAGA202412} characterize the tops-only rules without obvious
manipulations and apply the characterization to generalized median voter
schemes and voting by committees; our results for coalitions of size one are
theirs.
Outside voting, non-obvious manipulability has been studied in cake-cutting
\citep{ortega2022obvious}, the allocation of indivisible goods
\citep{psomas2022fair}, assignment problems \citep{troyan2024nonobvious},
allotment and division problems
\citep{arribillaga2025,ARRIBILLAGA2025112344,arribillaga2025division},
two-sided matching \citep{arribillaga2025matching}, and budget-feasible
mechanism design \citep{dekeijzer2026breaking}.

Our paper is closely related to two contemporaneous contributions,
\citet{doi2026obvious} and \citet{demoor2026nonobvious}, which study
the role of groups in non-obvious manipulability in Shapley--Scarf
housing markets. These contributions and ours pursue a common
research agenda through independently developed approaches.
Both papers cite an earlier version of our
project,\footnote{The version cited by both papers,
\citet{bonifacio2026obvious}, used a different definition, in which
the best and worst outcomes of an option set were evaluated
separately for each coalition member. The joint definition adopted
here, and the results that depend on it, were obtained independently
of \citet{doi2026obvious}.}
and their connections with the present paper concern both the
definition of group extensions and their implications for
mechanism design.

The connection with \citet{doi2026obvious} is particularly close:
they independently introduce the same notion of obvious group
manipulation studied here. They call the resulting property NOGM
and show that, together with individual rationality and pair
efficiency, it characterizes the top trading cycles mechanism.
Their formulation uses Pareto-undominated best and worst outcomes;
under our assumption of strict preferences over alternatives,
the two formulations coincide
(Remark~\ref{rem:formulations}).
They also establish a characterization of obvious group
manipulations in a general mechanism-design framework that,
when specialized to our voting setting, implies our independently
obtained Proposition~\ref{prop:indistinguishability} and also
covers the case of two alternatives.

\citet{demoor2026nonobvious} pursue a different extension.
Agents are partitioned into exogenously given groups, and each
agent knows the reports of the other members of her group but
not those of outsiders. Deviations remain individual: groups
determine the information available to a deviating agent,
whereas in our framework they identify the agents who can
coordinate their reports and evaluate a deviation jointly.
Their approach therefore explores the informational role
of groups, while ours studies coordinated manipulation.

A common motivation underlies these contemporaneous developments:
individual non-obvious manipulability may impose only weak
restrictions on mechanisms. Both papers start from this
observation in housing markets. Its counterpart in voting,
where many manipulable rules admit no obvious individual
manipulation, motivates the present paper.
The three contributions thus investigate complementary ways
of incorporating groups into the analysis of obvious
manipulations, through richer information or coordinated
deviations, and examine the resulting restrictions in
different economic environments.

On the universal domain, strategy-proofness and group strategy-proofness both
reduce to dictatorship when there are at least three alternatives.  On
restricted domains they need not coincide, and
\citet{lebreton2009equivalence} and \citet{BARBERA20101648} give conditions
on the domain under which they do.  The vulnerability of voting rules to
coalitions has also been measured by the frequency of profiles at which some
coalition can manipulate \citep{lepelley1987proportion} and by the size of
the coalitions needed to manipulate
\citep{slinko2004how,pritchardWilson2007}.
\citet{kondrratev2020measuringmajoritypower} measure the smallest share of
the voters that can guarantee the election of one of its most preferred
candidates, or veto its least preferred ones, whatever the other voters do.
Our
thresholds answer a different question: the smallest size at which some
coalition has an \emph{obvious} manipulation somewhere on the domain.

Finally, manipulation under limited information about the other voters has
been modeled by specifying the set of profiles that a manipulator considers
possible.  \citet{conitzerWalshXia2011} ask whether some vote does at least
as well as the truthful one at each such profile and strictly better at some
profile, and \citet{reijngoud2012voter} and \citet{endriss2016strategic}
study how voters respond to poll information and to other forms of partial
information.  Earlier, \citet{moulin1981prudence} and
\citet{BARBERA198249} studied voters who evaluate their strategies by the
worst outcomes that they can lead to.  Here a coalition knows nothing about
the outsiders' votes, but it knows the range of each of its reports.

\paragraph{Organization.} Section~\ref{sec:model} introduces the model and
group strategy-proofness. Section~\ref{sec:obvious} defines obvious group
manipulations and establishes their informational foundation.
Section~\ref{sec:binary} treats the case of two alternatives.
Section~\ref{sec:limits} develops the impossibility results for
tops-only rules and rules satisfying the majority criterion, and
then compares the coalition sizes needed for obvious manipulation.
Section~\ref{sec:possibility} studies the exceptional case of four
voters and three alternatives.
Section~\ref{sec:applications} applies the tops-only characterization to
generalized median voter schemes and voting by committees.
Section~\ref{sec:conclusion} concludes.
Appendix~\ref{app:n4m3} contains the proof of Theorem~\ref{thm:n4m3}.

\section{Preliminaries}\label{sec:model}

Let $N=\{1,\ldots,n\}$ be the set of voters and let $\alt$ be the set of
alternatives, with $|\alt|=m$ and $n,m\geq2$.  Each voter $i$ has a
preference $P_i$ in the set $\linalt$ of all strict linear orders on $\alt$.
We write $x\,P_i\,y$ if $i$ prefers $x$ to $y$, and $t(P_i)$ for her most
preferred alternative, her \emph{top}.  A profile is $P=(P_i)_{i\in N}$.  For
a nonempty coalition $C\subseteq N$, we write $P_C=(P_i)_{i\in C}$,
$P_{-C}=P_{N\setminus C}$ and $P=(P_C,P_{-C})$; the voters in $N\setminus C$
are the \emph{outsiders}.  A coalition is \emph{proper} if it is nonempty and
different from $N$.

A \emph{voting rule} $f:\linalt^N\to\alt$ selects an alternative at each
profile.  Throughout, rules are onto.  A rule is \emph{efficient} if there
are no profile $P$ and alternative $x$ such that $x\,P_i\,f(P)$ for every
$i\in N$.  It is \emph{dictatorial} if there is a voter $d$ such that
$f(P)=t(P_d)$ for every $P$.

Voter $i$ can \emph{manipulate} $f$ at $P$ if some report $P'_i$ gives
$f(P'_i,P_{-i})\,P_i\,f(P)$.  A rule is \emph{strategy-proof} if no voter can
manipulate it at any profile.  The classical extension allows a coalition to
choose its reports jointly.  A report $P'_C\in\linalt^C$ is a \emph{group
manipulation} for $C$ at $P$ if
\[
 f(P'_C,P_{-C})\,P_i\,f(P)\qquad\text{for every }i\in C.
\]
Every member must gain, although some members may keep their truthful
ballots: a coalition commits to a joint report, and a member's commitment to
her truthful ballot is part of that report.  A rule is \emph{group
strategy-proof} if it admits no group manipulation.

Group strategy-proofness implies strategy-proofness.  Whether
strategy-proofness already protects against coalitions depends on the
preference domain.  On the domain of single-peaked preferences, for instance,
strategy-proofness implies group strategy-proofness, and
\citet{lebreton2009equivalence} and \citet{BARBERA20101648} give conditions
on the domain under which the two requirements coincide.  On the universal
domain considered here, the Gibbard--Satterthwaite theorem leaves only
dictatorships once there are at least three alternatives
\citep{gibbard1973manipulation,satterthwaite1975strategy}, and every
dictatorship is group strategy-proof: a coalition that changes the outcome
must include the dictator, who already obtains her top.

For non-dictatorial rules, the question is then which opportunities to
manipulate can be recognized.  \citet{TROYAN2020104970} propose
\emph{non-obvious manipulability}: a voter must be unable to improve either
her worst possible outcome or her best possible outcome by misreporting.  We
ask the corresponding question for a group.  If its members know one
another's preferences but do not know how the outsiders will vote, when can
they recognize a joint deviation from which all of them gain?

\section{Obvious group manipulations}\label{sec:obvious}

The \emph{option set} of a report $P_C$ of a nonempty coalition $C$ is the
set of outcomes that remain possible once the coalition submits that report:
\[
 O^f(P_C)=\{f(P_C,P_{-C}): P_{-C}\in\linalt^{N\setminus C}\}.
\]
We write $O(P_C)$ when the rule is understood. For a single voter this is
the usual option set $O^f(P_i)$; for the grand coalition it is the singleton
$\{f(P)\}$. A report $P_C$ \emph{vetoes} $x$ if $x\notin O(P_C)$, and it
\emph{forces} $x$ if $O(P_C)=\{x\}$; a coalition \emph{can veto} (\emph{can
force}) $x$ if one of its reports does.

Compare the option sets of the truthful report and of a deviation. The
deviation is recognizably profitable if it eliminates an outcome that all
members rank below every outcome that is still possible, or if it makes
attainable an outcome that all members rank above every outcome that is
possible under truthful reporting.

\begin{definition}\label{def:obvious}
Let $C\subseteq N$ be nonempty, let $P_C\in\linalt^C$ be the true preferences
of its members, let $P'_C\in\linalt^C$ be a report, and write $T=O(P_C)$ and
$F=O(P'_C)$. The report $P'_C$ is an \emph{obvious group manipulation} for
$C$ at $P_C$ if at least one of the following holds:
\begin{enumerate}[label=(\textup{W}),leftmargin=3em]
\item there is $x\in T$ such that $y\,P_i\,x$ for every $y\in F$ and every
      $i\in C$;
\end{enumerate}
\begin{enumerate}[label=(\textup{B}),leftmargin=3em]
\item there is $y\in F$ such that $y\,P_i\,x$ for every $x\in T$ and every
      $i\in C$.
\end{enumerate}
A rule is \emph{not obviously manipulable by groups} (NOMG) if no coalition
has an obvious group manipulation, and \emph{obviously manipulable by groups}
(OMG) otherwise.
\end{definition}

Condition (W) says that some outcome that the coalition may obtain by telling
the truth is unanimously worse than anything it may obtain by misreporting;
condition (B) says that some outcome that it may obtain by misreporting is
unanimously better than anything it may obtain by telling the truth. The
definition only uses the two option sets and the members' preferences.
As discussed in the Introduction, it coincides with the definition of
\citet{doi2026obvious}, as clarified in the following remark.

\begin{remark}[Two formulations]\label{rem:formulations}
\citet{doi2026obvious} state the definition differently. For a set of
outcomes $O$, let $B_C(O)$ be the set of elements of $O$ that are not Pareto
dominated, for the members of $C$, by another element of $O$, and let
$W_C(O)$ be the set of elements of $O$ that do not Pareto dominate another
element of $O$. In their definition, $P'_C$ is an obvious group manipulation
at $P_C$ if some element of $B_C(F)$ Pareto dominates every element of
$B_C(T)$, or some element of $W_C(T)$ is Pareto dominated by every element of
$W_C(F)$. Such a formulation is needed when members can be indifferent
between distinct outcomes, as in allocation problems, where an agent only
cares about her own assignment. Here preferences over outcomes are strict,
so $y$ Pareto dominates $x\neq y$ for $C$ if and only if $y\,P_i\,x$ for
every $i\in C$. This relation is transitive, and therefore every element of
a finite set $O$ is equal to, or dominated by, an element of $B_C(O)$, and is
equal to, or dominates, an element of $W_C(O)$. It follows that their two
conditions are equivalent to (B) and (W), respectively. For instance, if
$y\in F$ dominates every element of $T$, then so does an element of $B_C(F)$
that is equal to or dominates $y$; conversely, if $y\in B_C(F)$ dominates
every element of $B_C(T)$, then it dominates every element of $T$, so that
(B) holds with $y$. The argument for (W) is symmetric. Hence NOMG is the
property that \citet{doi2026obvious} call NOGM, specialized to voting.
\end{remark}

At the two extremes of coalition size the conditions take a familiar form.
For a single voter they compare the worst and the best elements of the two
option sets; for the grand coalition each report determines one
outcome.\footnote{Parts (i) and (iii) of Lemma~\ref{lem:basic} have
counterparts for housing markets in \citet{doi2026obvious}, who show that
group strategy-proofness implies their property and that an onto mechanism
with that property is Pareto efficient.}

\begin{lemma}\label{lem:basic}
Let $C$, $P_C$, $P'_C$, $T$ and $F$ be as in Definition~\ref{def:obvious}.
\begin{enumerate}[label=(\roman*)]
\item If $P'_C$ is an obvious group manipulation for $C$ at $P_C$, then there
      is $P_{-C}$ such that $P'_C$ is a group manipulation for $C$ at
      $(P_C,P_{-C})$. Hence every group strategy-proof rule is NOMG.
\item For $C=\{i\}$, $P'_i$ is an obvious group manipulation at $P_i$ if and
      only if either the worst element of $F$ is preferred to the worst
      element of $T$, or the best element of $F$ is preferred to the best
      element of $T$, where worst and best refer to $P_i$. Thus the absence
      of obvious manipulations by single voters is non-obvious manipulability
      in the sense of \citet{TROYAN2020104970}.
\item For $C=N$, $P'_N$ is an obvious group manipulation at $P$ if and only
      if it is a group manipulation at $P$. Consequently, the grand
      coalition has no obvious group manipulation if and only if the rule is
      efficient, and a rule is NOMG if and only if it is efficient and no
      proper coalition has an obvious group manipulation.
\end{enumerate}
\end{lemma}

\begin{proof}
(i) If (W) holds with $x$, choose $P_{-C}$ with $f(P_C,P_{-C})=x$; then
$f(P'_C,P_{-C})\in F$ is strictly preferred to $x$ by every member. If (B)
holds with $y$, choose $P_{-C}$ with $f(P'_C,P_{-C})=y$; then
$f(P_C,P_{-C})\in T$ is strictly worse than $y$ for every member.

(ii) If some $x\in T$ satisfies (W), then so does the worst element of $T$.
Hence (W) holds if and only if every element of $F$, that is, the worst
element of $F$, is preferred to the worst element of $T$. In the same way,
(B) holds if and only if the best element of $F$ is preferred to the best
element of $T$.

(iii) For $C=N$ both option sets are singletons, and (W) and (B) both say
that $f(P')\,P_i\,f(P)$ for every $i$. If $f$ is not efficient, say
$x\,P_i\,f(P)$ for every $i$, then since $f$ is onto there is $P'$ with
$f(P')=x$, a group manipulation by $N$. Conversely, a group manipulation by
$N$ exhibits a violation of efficiency.
\end{proof}

These properties connect the definition with individual non-obvious
manipulability and group strategy-proofness. We next establish its
informational foundation. In \citet{TROYAN2020104970}, an agent who
cannot reason contingently about the reports of others evaluates a
report through the best and worst outcomes it may produce. Our
coalition faces the same limitation with respect to outsiders, while
its members know each other's preferences and can commit to a joint
report, as in the definition of group strategy-proofness. The
comparison must therefore be decisive for the coalition as a whole:
either a truthful outcome is unanimously worse than every possible
outcome of the deviation, or an outcome of the deviation is
unanimously better than every possible truthful outcome.

Knowing these option sets does not reveal how the outcomes of two
reports are paired at the same outsider profile. To make this
precise, say that rules $f$ and $g$ are
\emph{$C$-indistinguishable} if $O^f(P_C)=O^g(P_C)$ for every
$P_C\in\linalt^C$. Such rules have the same range, the union of
the option sets of $C$, so that $g$ is onto whenever $f$ is.

The next result shows that the definition captures exactly those
deviations that the coalition can identify as group manipulations
from this information alone: a misreport is an obvious group
manipulation if and only if it remains a group manipulation under
every $C$-indistinguishable voting rule. It thus extends to
coalitions the characterization of obvious manipulations in
\citet{TROYAN2020104970}.\footnote{Our Proposition~1, formulated in terms of option sets, is equivalent
to Theorem~3 of \citet{doi2026obvious}, stated for general mechanisms,
when specialized to voting. Their result also covers the case of two
alternatives. The proofs were obtained independently and use similar
arguments, building on the construction of \citet{TROYAN2020104970}.}

\begin{proposition}\label{prop:indistinguishability}
Let $m\geq3$, let $C\subseteq N$ be nonempty and let
$P_C,P'_C\in\linalt^C$. Then $P'_C$
is an obvious group manipulation for $C$ at $P_C$ under $f$ if and only if,
for every rule $g$ that is $C$-indistinguishable from $f$, there is an
outsider profile $Q_{-C}\in\linalt^{N\setminus C}$ such that
\[
 g(P'_C,Q_{-C})\;P_i\;g(P_C,Q_{-C})\qquad\text{for every }i\in C.
\]
\end{proposition}

\begin{proof}
Write $T=O^f(P_C)$ and $F=O^f(P'_C)$. If $C=N$, the only
$N$-indistinguishable rule is $f$ itself and both statements say that every
member prefers $f(P')$ to $f(P)$ (Lemma~\ref{lem:basic}). Assume $C\neq N$.

\emph{Only if.} Let $g$ be $C$-indistinguishable from $f$. If (W) holds with
$x\in T=O^g(P_C)$, choose $Q_{-C}$ with $g(P_C,Q_{-C})=x$; then
$g(P'_C,Q_{-C})\in F$ is unanimously preferred to $x$. If (B) holds with
$y\in F=O^g(P'_C)$, choose $Q_{-C}$ with $g(P'_C,Q_{-C})=y$; then
$g(P_C,Q_{-C})\in T$ is unanimously worse than $y$.

\emph{If.} Suppose neither (W) nor (B) holds. If $P'_C=P_C$, take $g=f$.
Otherwise, call a pair $(x,y)\in T\times F$ \emph{improving} if
$y\,P_i\,x$ for every $i\in C$. Since (W) fails, for every $x\in T$ there is
$\varphi(x)\in F$ such that $(x,\varphi(x))$ is not improving; since (B)
fails, for every $y\in F$ there is $\psi(y)\in T$ such that $(\psi(y),y)$ is
not improving. Since $m\geq3$, the number of outsider profiles is
$(m!)^{|N\setminus C|}\geq m!\geq2m\geq|T|+|F|$. Choose distinct
outsider profiles $Q^x$ ($x\in T$) and $\widehat Q^y$ ($y\in F$), and
define $g$ by
\[
 g(P_C,Q^x)=x,\quad g(P'_C,Q^x)=\varphi(x),\qquad
 g(P_C,\widehat Q^y)=\psi(y),\quad g(P'_C,\widehat Q^y)=y,
\]
by $g(P_C,Q)=x_0$ and $g(P'_C,Q)=\varphi(x_0)$ at every remaining outsider
profile $Q$, for a fixed $x_0\in T$, and by $g=f$ at all profiles at which $C$
reports neither $P_C$ nor $P'_C$. Then $O^g(P_C)=T$, $O^g(P'_C)=F$, and all other option sets of $C$ are
unchanged, so $g$ is $C$-indistinguishable from $f$. By construction,
no outsider profile yields an improving pair.
\end{proof}

The outsider profile at which the coalition gains may depend on
the rule, but for each rule the same profile must work for every
member. Conversely, if a deviation is not obvious, some
$C$-indistinguishable rule makes it unprofitable for the coalition
at every outsider profile. Option sets alone therefore suffice
to identify exactly the obvious group manipulations.

\section{Two alternatives}\label{sec:binary}

Let $\alt=\{a,b\}$.  A vote for $a$ represents the preference $a\,P_i\,b$,
and $a^C$ denotes the report of coalition $C$ at which every member votes for
$a$.  With two alternatives, a rule is efficient if and only if it selects
$a$ whenever all voters vote for $a$, for each $a\in\alt$.

\begin{proposition}\label{prop:binary}
Let $m=2$ and let $C$ be a nonempty coalition.  Then $C$ has no obvious group
manipulation if and only if, for each alternative $a$:
\begin{enumerate}[label=(\roman*)]
\item if some report of $C$ forces $a$, then $a^C$ forces $a$; and
\item $a\in O(a^C)$.
\end{enumerate}
In particular, a rule is NOMG if and only if it is efficient and every proper
coalition that can force an alternative can force it by voting unanimously
for it.
\end{proposition}

\begin{proof}
If two members of $C$ disagree, neither outcome is strictly better than the
other for every member, so neither (W) nor (B) can hold.  Suppose that all
members prefer $a$ to the other alternative, $b$, write $T=O(a^C)$, and let
$F$ be the option set of a report of $C$.  Condition (W) holds if and only if
$b\in T$ and $F=\{a\}$; ruling this out for every report gives (i).
Condition (B) holds if and only if $T=\{b\}$ and $a\in F$.  Since the rule is
onto, some report of $C$ leaves $a$ open, so ruling (B) out for every report
gives (ii).

For the last statement, NOMG implies efficiency by
Lemma~\ref{lem:basic}(iii).  Conversely, let the rule be efficient.  Letting
all outsiders vote for $a$ shows that (ii) holds for every coalition, and (i)
holds for $C=N$ because $a^N$ forces $a$.  Hence the rule is NOMG if and only
if (i) holds for every proper coalition.
\end{proof}

Condition (i) says that the power to force an alternative never requires a
coalition to split its votes.  It is much weaker than strategy-proofness.

\begin{example}[Parity rules]\label{ex:parity}
Let $\alt=\{x,y\}$, let $n\geq3$ be odd, and let $f$ select $y$ if and only
if an odd number of voters vote for $y$.  For every proper coalition $C$ and
every report of $C$, a single outsider can change the parity, so every option
set of $C$ is $\alt$ and $C$ forces nothing.  Since $n$ is odd, $f$ selects
$y$ when all voters vote for $y$ and $x$ when all voters vote for $x$, so $f$
is efficient.  By Proposition~\ref{prop:binary}, $f$ is NOMG.  It is not
strategy-proof: if voter $i$, who prefers $y$, and exactly one other voter
vote for $y$, the outcome is $x$, and if $i$ switches to $x$ the outcome
becomes $y$.
\end{example}

Conversely, efficiency alone leaves open the possibility that a coalition can
force its preferred alternative only by splitting its votes.

\begin{example}[Efficiency is not enough]\label{ex:binary-eff}
Let $n=3$ and $\alt=\{x,y\}$, and let $f$ select $y$ if voter~$1$ votes for
$y$ and voter~$2$ votes for $x$, and the alternative for which voter~$3$
votes otherwise.  This rule is efficient.  The coalition $C=\{1,2\}$ forces
$y$ with the report at which voter~$1$ votes for $y$ and voter~$2$ votes for
$x$, but $O(y^C)=\alt$ because voter~$3$ decides, so (i) fails.  Concretely,
let voters $1$ and $2$ prefer $y$.  If voter~$2$ switches to $x$ while
voter~$1$ keeps her vote, the outcome is $y$ whatever voter~$3$ does.  Since
$x$ belongs to the truthful option set and both members prefer $y$, the only
element of the deviating option set, this is an obvious group manipulation of
type~(W).
\end{example}

\section{Impossibility results and coalition size}\label{sec:limits}

We study two overlapping classes of voting rules. Tops-onliness
restricts the information used to choose an outcome, whereas the
majority criterion guarantees that a strict majority can impose
its common top alternative. Thus, the two classes capture distinct
aspects of collective choice rather than partitioning the set of
voting rules. We first establish the restrictions that NOMG imposes
within each class and then compare the coalition sizes responsible
for them. Throughout, preferences range over the universal domain.

\subsection{Tops-only rules}\label{subsec:tops}
The characterization below holds for every $m\geq2$. For the
equivalence with group strategy-proofness and the dictatorship
result, we assume $m\geq3$.

A rule is \emph{tops-only} if $f(P)=f(P')$ whenever $t(P_i)=t(P'_i)$ for
every $i\in N$.  Tops-only rules include plurality, generalized median voter
schemes and voting by committees; on many domains, strategy-proofness and
unanimity imply tops-onliness \citep{chatterji2011tops}.  For a tops-only
rule we use the same symbol for
the induced map $f:\alt^N\to\alt$; thus $f(\bt)$ is the outcome when the
vector of reported tops is $\bt=(t_1,\ldots,t_n)$, and $O(\br_C)$ is the
option set of a vector of tops $\br_C\in\alt^C$.  For $\bt_C\in\alt^C$ let
$\tau(\bt_C)=\{t_i:i\in C\}$ be the set of reported tops, for $x\in\alt$ let
$S_x(\bt)=\{i\in N:t_i=x\}$ be the \emph{$x$-support} of $\bt$, and let $x^C$
denote the report of $C$ at which every member reports $x$.  A tops-only rule
is efficient if and only if $f(\bt)\in\tau(\bt)$ for every $\bt$: if the
outcome is a reported top it cannot be Pareto dominated, and if $f(\bt)$ is
not a reported top, there are preferences with the reported tops at which
every voter ranks some reported top $z$ above $f(\bt)$.

For a tops-only rule, the option sets of a coalition depend only on the tops
that it reports, while the lower ranks of its members' true preferences are
free.  We can therefore test whether a coalition has an obvious group
manipulation using only reported tops and option sets.

\begin{lemma}\label{lem:tops-reform}
Let $f$ be tops-only and let $C\subseteq N$ be nonempty.  The coalition $C$
has an obvious group manipulation, for some true preferences of its members,
if and only if there are $\bt_C,\br_C\in\alt^C$ such that
\begin{enumerate}[label=(\textup{W}$'$),leftmargin=3.2em]
\item some $x\in O(\bt_C)\setminus O(\br_C)$ satisfies
      $x\notin\tau(\bt_C)$; or
\end{enumerate}
\begin{enumerate}[label=(\textup{B}$'$),leftmargin=3.2em]
\item $O(\bt_C)\cap\tau(\bt_C)=\varnothing$ and
      $O(\br_C)\not\subseteq O(\bt_C)$.
\end{enumerate}
\end{lemma}

\begin{proof}
Let $\bt_C$ be the tops of the true preferences, $\br_C$ the tops of the
report, $T=O(\bt_C)$ and $F=O(\br_C)$.  If (W) holds with $x$, then
$x\notin F$, and $x$ is not a member's top because every member ranks the
elements of $F$ above it; this is (W$'$).  Conversely, given (W$'$), let
every member rank her top first and $x$ last; then every element of $F$ is
above $x$ for every member.  If (B) holds with $y$, then $y\notin T$, and no
member's top is in $T$, since otherwise $y$ would have to be ranked above
that member's top; this is (B$'$).  Conversely, given (B$'$) and
$y\in F\setminus T$, let every member rank her top first, $y$ next (if it is
different from her top), and the elements of $T$ below; since no top lies in
$T$, (B) holds.
\end{proof}

\begin{definition}\label{def:cv}
Let $f$ be tops-only.  A nonempty coalition $C$ satisfies the
\emph{conditional veto} if, whenever some report of $C$ vetoes an alternative
$x$, every report $\bt_C$ with $x\notin\tau(\bt_C)$ vetoes $x$.
\end{definition}

The conditional veto is exactly what prevents (W$'$).  To prevent (B$'$), a
report must leave open at least one of the tops that it contains.

\begin{theorem}\label{thm:characterization}
Let $f$ be tops-only and let $C$ be a nonempty coalition.  Then $C$ has no
obvious group manipulation if and only if
\begin{enumerate}[label=(\roman*)]
\item $C$ satisfies the conditional veto; and
\item every report of $C$ leaves open the top of at least one of its
      members: $O(\bt_C)\cap\tau(\bt_C)\neq\varnothing$ for every
      $\bt_C\in\alt^C$.
\end{enumerate}
If $f$ is efficient, then (ii) holds for every coalition and (i) holds for
the grand coalition.  Hence an efficient tops-only rule is NOMG if and only
if every proper coalition satisfies the conditional veto.
\end{theorem}

\begin{proof}
By Lemma~\ref{lem:tops-reform}, $C$ has no obvious group manipulation if and
only if (W$'$) and (B$'$) fail for all $\bt_C,\br_C$.  The failure of (W$'$)
for all $\bt_C,\br_C$ is the conditional veto.  For (B$'$), note that the
union of the option sets of $C$ is $\alt$, because $f$ is onto.  If
$O(\bt_C)$ contains none of the tops in $\bt_C$, choose $y\in\tau(\bt_C)$ and
a report $\br_C$ with $y\in O(\br_C)$; then $y\notin O(\bt_C)$, so (B$'$)
holds.  Conversely, if (ii) holds, then (B$'$) fails for every pair of
reports.

Let $f$ be efficient, so that $f(\bt)\in\tau(\bt)$ for every $\bt$.  For a
proper coalition $C$ and a report $\bt_C$, let the outsiders report the top
of some member of $C$; the outcome lies in $\tau(\bt_C)$, so (ii) holds.  For
$C=N$, $O(\bt)=\{f(\bt)\}\subseteq\tau(\bt)$, so (ii) holds and no
alternative outside $\tau(\bt)$ is left open, which gives (i).  The last
statement follows from Lemma~\ref{lem:basic}(iii).
\end{proof}

For a single voter, (i) implies (ii).  Indeed, if reporting $x$ vetoed $x$,
then by the conditional veto every report of that voter would veto $x$, which
contradicts ontoness.  For $C=\{i\}$, Theorem~\ref{thm:characterization} is
therefore the characterization of \citet{ARRIBILLAGA202412}: a tops-only rule
has no obvious individual manipulation if and only if, whenever a voter can
veto an alternative, she vetoes it with every preference that does not have
that alternative on top.

Applying the characterization to all coalitions has a strong implication.  If
a tops-only rule protects every coalition, then the voters who support the
winner at any given profile include a group that can secure that outcome by
itself.

\begin{theorem}\label{thm:gsp}
Let $m\geq3$ and let $f$ be an onto tops-only rule.  The following are
equivalent:
\begin{enumerate}[label=(\roman*)]
\item $f$ is NOMG;
\item $f$ is efficient and every proper coalition satisfies the conditional
      veto;
\item $f$ is group strategy-proof;
\item $f$ is dictatorial.
\end{enumerate}
\end{theorem}

\begin{proof}
(i)$\Longleftrightarrow$(ii).  NOMG implies efficiency by
Lemma~\ref{lem:basic}(iii), and for efficient rules the equivalence is the
last statement of Theorem~\ref{thm:characterization}.

(ii)$\Longrightarrow$(iii).  We proceed in four steps.

\emph{Step 1: winning supports contain kernels.}  Let $f(\bt)=a$.  Among the
profiles of tops at which $a$ is selected and whose $a$-support is contained
in $S_a(\bt)$, choose one, $\bt^{\ast}$, whose $a$-support $K$ is minimal
with respect to inclusion; $K$ is nonempty by efficiency.  We claim that
\begin{equation}\label{eq:kernel}
 f(a^K,\br_{-K})=a\qquad\text{for every }\br_{-K}\text{ with }
 a\notin\tau(\br_{-K}).
\end{equation}
If $K=N$, the claim is that $a$ is selected when all voters report $a$, which
holds by efficiency.  Otherwise, $C=N\setminus K$ is a proper coalition whose
report at $\bt^{\ast}$ avoids $a$ and leaves $a$ open.  By the conditional
veto, no report of $C$ vetoes $a$.  Let $\br_C$ avoid $a$ and choose $\br_K$
such that $f(\br_K,\br_C)=a$.  The $a$-support of $(\br_K,\br_C)$ is contained
in $K\subseteq S_a(\bt)$, so by minimality it equals $K$; that is,
$\br_K=a^K$.

\emph{Step 2: winning supports of different alternatives intersect.}  Let
$f(\bt)=x$ and $f(\bt')=z$ with $x\neq z$, and suppose that $S_x(\bt)$ and
$S_z(\bt')$ are disjoint.  Let $K\subseteq S_x(\bt)$ and
$L\subseteq S_z(\bt')$ be given by Step~1, and choose
$c\in\alt\setminus\{x,z\}$, which is possible because $m\geq3$.  Consider the
profile at which the members of $K$ report $x$, the members of $L$ report
$z$, and every other voter reports $c$.  The voters outside $K$ avoid $x$, so
the outcome is $x$ by \eqref{eq:kernel}; the voters outside $L$ avoid $z$, so
the outcome is $z$; a contradiction.

\emph{Step 3: kernels force their alternative.}  Let $f(\bt)=a$ and let $K$
be given by Step~1.  We claim that $f(a^K,\br_{-K})=a$ for every $\br_{-K}$.
If $K=N$ there is nothing to prove.  Otherwise, suppose that
$z=f(a^K,\br_{-K})\neq a$ for some $\br_{-K}$.  The $z$-support
of $(a^K,\br_{-K})$ is contained in $N\setminus K$, whereas for any
$\br'_{-K}$ that avoids $a$, the alternative $a$ is selected at
$(a^K,\br'_{-K})$ with $a$-support $K$.  This contradicts Step~2.

\emph{Step 4.}  Suppose that a coalition $C$ has a group manipulation at a
profile with tops $\bt$, which changes the outcome from $a=f(\bt)$ to some
$y$.  Every member strictly prefers $y$ to $a$, so no member has top $a$;
that is, $C\cap S_a(\bt)=\varnothing$.  Let $K\subseteq S_a(\bt)$ be given by
Step~1.  The members of $K$ keep reporting $a$ after the deviation, so by
Step~3 the outcome is still $a$, a contradiction.

(iii)$\Longrightarrow$(iv).  A group strategy-proof rule is strategy-proof.  Since
$f$ is onto and $m\geq3$, the Gibbard--Satterthwaite theorem implies that $f$
is dictatorial.

(iv)$\Longrightarrow$(i).  A dictatorial rule is group strategy-proof, and group
strategy-proof rules are NOMG by Lemma~\ref{lem:basic}(i).
\end{proof}

Theorem~\ref{thm:gsp} says that, for tops-only rules, restricting coalitions
to the deviations that they can recognize from option sets does not enlarge
the set of admissible rules beyond the group strategy-proof ones.  The third
alternative is essential: with two alternatives every rule is tops-only, and
the parity rules of Example~\ref{ex:parity} (in Section~\ref{sec:binary}) are NOMG but not strategy-proof.
The contrast with the individual case is also sharp.  Fix a status quo
$s\in\alt$ and let $f(\bt)=a$ if every voter reports $a$, and $f(\bt)=s$
otherwise.  This rule is onto, tops-only and non-dictatorial.  A voter who
reports $a$ has option set $\{a,s\}$, which contains her reported top.  She
can veto every alternative other than $s$, and every report that does not
have such an alternative on top vetoes it.  By
Theorem~\ref{thm:characterization}, the rule has no obvious individual
manipulation.  It is OMG because it is not efficient
(Lemma~\ref{lem:basic}(iii)).

\subsection{Rules satisfying the majority criterion}\label{subsec:majority}

We now allow voting rules to depend on the entire rankings.
The next impossibility follows from the majority criterion and
therefore applies, in particular, to Condorcet-consistent rules.

A rule satisfies the \emph{majority criterion} if it selects $x$ whenever
more than $n/2$ voters rank $x$ first; a coalition with more than $n/2$
members can then force any alternative by ranking it first, even under a rule
that uses the entire rankings.  A rule is \emph{Condorcet consistent} if it
selects the Condorcet winner whenever one exists; an alternative ranked first
by more than $n/2$ voters is a Condorcet winner, so Condorcet consistency
implies the majority criterion.  \citet{dasguptaMaskin2008,dasguptaMaskin2020}
single out majority rule by its performance on restricted domains; in the
second paper the axioms include a strategy-proofness requirement that covers
manipulations by coalitions.  On the
universal domain, the majority criterion alone creates obvious group
manipulations: if the members of a majority all prefer some alternative to
the alternative that is selected, the selected alternative itself is a
witness for condition (W).

The construction below is related to the argument of \citet{nakamura1979}
for the emptiness of the core of a simple game with ordinal preferences.  A
cycle suffices if each alternative in the cycle can be replaced by a
coalition that prefers, and can force, another alternative.

\begin{proposition}\label{prop:cyclic}
Let $m\geq3$, let $a,b,c$ be distinct alternatives, and suppose that $N$ can
be partitioned into nonempty sets $C_1,C_2,C_3$ such that every union of two
of them can force each of $a$, $b$ and $c$.  Then $f$ is OMG.
\end{proposition}

\begin{proof}
Consider the profile $P$ at which the members of $C_1$, $C_2$ and $C_3$ rank
$a,b,c$ in the orders $a\,b\,c$, $b\,c\,a$ and $c\,a\,b$, respectively, and
all other alternatives below these three in a common order.  Let $v=f(P)$.
If $v\notin\{a,b,c\}$, every voter ranks $a$ above $v$, and $C_1\cup C_2$ can
force $a$ while $v\in O(P_{C_1\cup C_2})$, so (W) holds with $v$.  If $v=a$,
every member of $C_2\cup C_3$ ranks $c$ above $a$; if $v=b$, every member of
$C_1\cup C_3$ ranks $a$ above $b$; if $v=c$, every member of $C_1\cup C_2$
ranks $b$ above $c$.  In each case that union can force the alternative that
it prefers, and (W) holds with $v$.
\end{proof}

\begin{theorem}\label{thm:majority-omg}
Let $n,m\geq3$ with $(n,m)\neq(4,3)$.  Every rule satisfying the majority
criterion is OMG.  In particular, every Condorcet-consistent rule is OMG.
\end{theorem}

\begin{proof}
If $n\neq4$, partition $N$ into three nonempty sets of sizes as equal as
possible; each has fewer than $n/2$ members (this fails only for $n=4$), so
the union of any two has more than $n/2$ members and
Proposition~\ref{prop:cyclic} applies.  If $n=4$, then $m\geq4$; consider the
profile
\[
 a\,b\,c\,d,\qquad b\,c\,d\,a,\qquad c\,d\,a\,b,\qquad d\,a\,b\,c,
\]
with all other alternatives below these four in a common order.  Each
$v\in\{a,b,c,d\}$ is ranked below its cyclic predecessor by three voters, and
any alternative outside $\{a,b,c,d\}$ is ranked below $a$ by all four.  Three
voters can force the alternative that they prefer by ranking it first, so (W)
holds with the alternative $v$ selected at this profile.
\end{proof}

Thresholds on the number of alternatives also appear in other impossibility
results for Condorcet-consistent rules: \citet{MOULIN198853} shows that every
such rule generates the no-show paradox when there are four or more
alternatives.  Here the only exception is a single electorate.

\subsection{How large must coalitions be?}\label{subsec:sizes}

The preceding results impose two restrictions on full immunity:
tops-only rules must be dictatorial when $m\geq3$, while rules
satisfying the majority criterion cannot be NOMG when $n,m\geq3$
and $(n,m)\neq(4,3)$. We now ask how large a coalition must be
before an obvious manipulation becomes possible. This question
applies to both classes. For $k\in\{1,\ldots,n\}$, a rule is \emph{$k$-NOMG} if no
coalition with at most $k$ members has an obvious group manipulation.  Let
$K(f)$ denote the largest such $k$, with $K(f)=0$ if some single voter has an
obvious manipulation.  Thus $K(f)=n$ if and only if $f$ is NOMG, and
otherwise $K(f)+1$ is the size of the smallest coalitions that can manipulate
obviously.  By Theorem~\ref{thm:characterization}, an efficient tops-only
rule is $k$-NOMG if and only if every proper coalition with at most $k$
members satisfies the conditional veto.

By Lemma~\ref{lem:basic}(iii), the grand coalition enters NOMG only through
efficiency, so for an efficient tops-only rule immunity to coalitions of up
to $n-1$ voters already implies dictatorship.  The next proposition shows that
neither efficiency nor the coalitions with $n-1$ members can be dispensed
with.  For $n\geq3$ and $d\in N$, the \emph{override rule} $f^{d}$ selects
$y$ if at least $n-1$ voters report $y$, and selects $t_d$ otherwise.  Since
$n-1>n/2$, at most one alternative is reported by $n-1$ voters, so $f^{d}$ is
well defined: voter $d$ decides unless all other voters agree on a different
alternative.  Now assign the values $0,\ldots,m-1$ to the $m$ different tops. The \emph{cyclic rule} selects
$\sum_{i\in N}t_i$ modulo $m$.

\begin{proposition}\label{prop:sizes}
Let $m\geq3$.
\begin{enumerate}[label=(\roman*)]
\item Every efficient, tops-only, $(n-1)$-NOMG rule is dictatorial.
\item For every $n\geq3$ and $d\in N$, the override rule $f^{d}$ is
      efficient, tops-only, non-dictatorial and $(n-2)$-NOMG.
\item For every $n\geq2$, the cyclic rule is tops-only, non-dictatorial and
      $(n-1)$-NOMG.
\end{enumerate}
Consequently, for efficient tops-only rules, $k$-NOMG implies dictatorship
if and only if $k\geq n-1$, and for tops-only rules in general if and only if
$k=n$.
\end{proposition}

\begin{proof}
(i) By Lemma~\ref{lem:basic}(iii) the rule is NOMG; apply
Theorem~\ref{thm:gsp}.

(ii) The outcome is always a reported top, so $f^{d}$ is efficient.  It is
not dictatorial.  Voter $d$ is not a dictator: if all voters other than $d$
report $y\neq t_d$, the outcome is $y$.  A voter $j\neq d$ is not a dictator
either: fix an alternative $a$ and a voter $l\notin\{d,j\}$, let $j$ and $l$
report two different alternatives other than $a$, and let all other voters
report $a$; no alternative is reported by $n-1$ voters, so the outcome is
$t_d=a\neq t_j$.  By Theorem~\ref{thm:characterization}, it remains to check
the conditional veto for every coalition $C$ with $c=|C|\leq n-2$ members.
There are $n-c\geq2$ outsiders.  For $\br_C\in\alt^C$ and $y\in\alt$, let
$\#_{\br_C}(y)$ be the number of members of $C$ who report $y$.

If $d\notin C$, let all outsiders, among them $d$, report $x$.  Every
$y\neq x$ is reported by at most $c\leq n-2$ voters, so the outcome is $x$.
Hence $O(\br_C)=\alt$ for every $\br_C$, and $C$ vetoes nothing.

If $d\in C$ and $d$ reports $a$, then
\[
 O(\br_C)=\{a\}\cup\{y\neq a:\#_{\br_C}(y)\geq c-1\}.
\]
Indeed, if all outsiders report $a$, no $y\neq a$ is reported by more than
$c-1\leq n-3$ voters, so the outcome is $a$; and $y\neq a$ can be selected
only if it is reported by $n-1$ voters, which the outsiders can achieve if and
only if $\#_{\br_C}(y)+(n-c)\geq n-1$.  If $c=1$, then $O(\br_C)=\alt$ for
every report and $C$ vetoes nothing.  If $c\geq2$ and $x\notin\tau(\bt_C)$,
then $x\neq t_d$ and $\#_{\bt_C}(x)=0<c-1$, so $\bt_C$ vetoes $x$.  In both
cases the conditional veto holds.

(iii) The rule is onto and tops-only.  It is not dictatorial: for every
voter $i$, if some other voter reports $1$ and all remaining voters report
$0$, the outcome is $t_i+1$ modulo $m$, which differs from $t_i$.  If $C$ is
a proper coalition and $j\notin C$, then for every report of the other voters,
$j$ can choose $t_j$ so that the sum takes any prescribed value modulo $m$.
Hence every option set of $C$ equals $\alt$, so $C$ satisfies both conditions
of Theorem~\ref{thm:characterization}.

For the last statement, combine (i)--(iii) with the fact that $k$-NOMG
implies $k'$-NOMG for $k'\leq k$, and with Theorem~\ref{thm:gsp} for $k=n$.
\end{proof}

The override rule shows where the proof of Theorem~\ref{thm:gsp} needs large
coalitions.  Let the members of $C=N\setminus\{d\}$ have true tops split
between two alternatives $b$ and $b'$, and let $a\notin\{b,b'\}$.  Truthful
reporting leaves $a$ open, because voter $d$ may report $a$, whereas the
report $b^C$ forces $b$; if all members rank $a$ last, this is an obvious
group manipulation of type (W).  Thus $K(f^d)=n-2$.  The cyclic rule has
$K(f)=n-1$; by part~(i), it is not efficient.

For rules satisfying the majority criterion, the relevant threshold
is instead determined by majority size. Coalitions with fewer than
half of the voters cannot restrict their option sets, since the
outsiders can force any alternative. Conversely, the profiles used
in Theorem~\ref{thm:majority-omg} allow a minimal strict majority
to manipulate obviously.

\begin{proposition}\label{prop:majority-sizes}
Let $n,m\geq3$ with $(n,m)\neq(4,3)$, and let $f$ satisfy the
majority criterion. Then no coalition with fewer than $n/2$
members has an obvious group manipulation, and some coalition
with exactly $\lfloor n/2\rfloor+1$ members has an obvious group
manipulation of type~(W). Consequently,
\[
K(f)=\frac{n-1}{2}\quad\text{if $n$ is odd},
\qquad
\frac n2-1\leq K(f)\leq\frac n2\quad\text{if $n$ is even}.
\]
In particular, these conclusions hold for every
Condorcet-consistent rule.
\end{proposition}

\begin{proof}
Let $C$ satisfy $|C|<n/2$. For every report $P_C$ and every
$x\in\alt$, the outsiders can rank $x$ first. Since they form a
strict majority, the outcome is $x$. Thus $O^f(P_C)=\alt$ for
every report, which rules out both (W) and (B).

For the upper bound, write $q=\lfloor n/2\rfloor+1$.
The construction in the proof of Theorem~\ref{thm:majority-omg}
provides a profile $P$, an alternative $y$, and a strict majority
$D$ such that $y\,P_i\,f(P)$ for every $i\in D$. Indeed, when
$n\neq4$, use the three blocks in that proof: if the selected
alternative belongs to the cycle, the union of the two blocks
that prefer its cyclic predecessor is such a majority; if it
lies outside the cycle, all voters prefer $a$ to it. When $n=4$
and $m\geq4$, the four-cycle construction gives three voters
who prefer the predecessor of the selected alternative, or all
four voters prefer $a$ if the outcome lies outside the cycle.

Choose $C\subseteq D$ with $|C|=q$ and let $P'_C$ be any report
at which every member ranks $y$ first. The original outsider
profile gives $f(P)\in O^f(P_C)$, whereas the majority criterion
gives $O^f(P'_C)=\{y\}$. Since every member of $C$ strictly
prefers $y$ to $f(P)$, condition (W) holds. The bounds on $K(f)$
follow, and coincide when $n$ is odd. Condorcet consistency
implies the majority criterion, so the same conclusions apply.
\end{proof}

For odd electorates, all rules satisfying the majority criterion
therefore have exactly the same protection threshold, regardless
of how they use the lower ranks. For even electorates, the only
unresolved size is $n/2$: the proposition does not determine
whether coalitions of that size can manipulate a given rule.
The exception $(n,m)=(4,3)$ is essential, as the next section
shows by exhibiting Condorcet-consistent Borda with $K(f)=n$.

Plurality belongs to both classes considered here: it is tops-only
and satisfies the majority criterion. Its exact threshold provides
a useful comparison with the preceding bounds.

Under plurality, what matters is how the size of the coalition compares with
the number of outsiders.  Let $f^{p}$ select, among the alternatives reported
as top by the largest number of voters, the highest one in a fixed strict
priority $\pri$ over $\alt$.

\begin{proposition}\label{prop:plurality}
Let $m\geq3$.  Plurality is $k$-NOMG if and only if $k<n/2$; that is,
$K(f^p)=\lceil n/2\rceil-1$.  Coalitions of fewer than half of the voters
have no obvious group manipulation, and some coalition with
$\lceil n/2\rceil$ members has one.
\end{proposition}

\begin{proof}
Plurality is tops-only and efficient.  Label the alternatives so that
$a_1\pri a_2\pri a_3\pri\cdots$.

If $|C|=c<n/2$, the outsiders outnumber the members of $C$; if they all
report $x$, then $x$ has more reports than any other alternative and wins.
Hence every option set of $C$ is $\alt$, $C$ vetoes nothing, and by
Theorem~\ref{thm:characterization} it has no obvious group manipulation.

Let $c=\lceil n/2\rceil$, which is at most $n-1$ because $n\geq2$, and let
$C$ be a coalition with $c$ members.  We exhibit a violation of the
conditional veto.

If $n$ is even, $c=n/2$ equals the number of outsiders.  Let every member
report $a_3$.  If all outsiders report $a_2$, then $a_2$ and $a_3$ tie with
$c$ reports each and $a_2$ wins, so $a_2\in O(a_3^C)$.  If instead every
member reports $a_1$, then $a_1$ has $c$ reports and no other alternative has
more, so $a_1$ wins every tie: $O(a_1^C)=\{a_1\}$, which vetoes $a_2$ although
$a_2\notin\tau(a_3^C)$.

If $n$ is odd, $c=(n+1)/2\geq2$ and there are $c-1$ outsiders.  Let
$\lceil c/2\rceil$ members report $a_2$ and $\lfloor c/2\rfloor$ members
report $a_3$.  If all outsiders report $a_1$, then $a_1$ has
$c-1\geq\lceil c/2\rceil$ reports and wins every tie, so $a_1$ belongs to the
option set of this report, although no member reports it.  If every member
reports $a_2$, then $a_2$ has $c$ reports, more than any other alternative can
obtain, so the option set is $\{a_2\}$, which vetoes $a_1$.
\end{proof}

Plurality and the override rule show how differently rules can behave in the
same electorate.  Plurality loses its protection as soon as a coalition can
match the outsiders' voting strength, whereas the override rule, which
concentrates power in one voter, protects against every coalition that
leaves at least two voters outside.

\section{A possibility result beyond tops-only rules}\label{sec:possibility}

The exception in Theorem~\ref{thm:majority-omg} and
Proposition~\ref{prop:majority-sizes} admits rules that are immune
to obvious manipulations by every coalition. With four voters and
three alternatives, the partition into three blocks
leaves one pair of blocks without a strict majority, and there is no fourth
alternative for the longer cycle.
Now we show that an entire family of scoring rules \citep{young1975} satisfying the majority criterion is NOMG.
A \emph{scoring rule} with vector
$s=(s_1,s_2,s_3)$, where $s_1\geq s_2\geq s_3$ and $s_1>s_3$, awards $s_j$
points to an alternative for each voter who ranks it in position $j$, and
selects, among the alternatives with the highest total score, the highest one
in a fixed strict priority $\pri$.  Since a positive affine transformation of
$s$ does not change the rule, we normalize $s=(1,t,0)$ with $t\in[0,1]$;
plurality is $t=0$ and Borda's rule is $t=\tfrac12$.  These rules are onto,
even for $t=1$: to elect $x$, let all voters rank it first and split their
second positions equally between the other two alternatives.  The result
below holds for every priority.

\begin{theorem}\label{thm:n4m3}
Let $n=4$ and $m=3$.  The scoring rule with vector $(1,t,0)$ is NOMG if and
only if $\tfrac12\leq t<\tfrac23$, that is, if and only if
$2s_2\geq s_1+s_3$ and $3s_2<2s_1+s_3$.
\end{theorem}

The proof is in Appendix~\ref{app:n4m3}. For $t<\tfrac12$, two voters with a common
bottom alternative can remove it from their option set by agreeing on their
top; for $\tfrac23\leq t<1$, three voters with a common preference can remove
their second-ranked alternative by having one of them demote it; and for
$t=1$ the rule is not efficient.  Every rule in the interval satisfies the
majority criterion: if three voters rank $x$ first, $x$ gets at least $3$
points and every other alternative at most $3t+1<3$.  Hence the exception in
Theorem~\ref{thm:majority-omg} cannot be removed.

\begin{corollary}\label{cor:borda}
With four voters and three alternatives, Borda's rule is Condorcet consistent
and NOMG.
\end{corollary}

\begin{proof}
Borda's rule is NOMG by Theorem~\ref{thm:n4m3} with $t=\tfrac12$.  For
Condorcet consistency, use the equivalent vector $(2,1,0)$: the score of an
alternative is the sum, over the other two alternatives, of the numbers of
voters who prefer it to that alternative.  A Condorcet winner is preferred to
each other alternative by at least three voters, so its score is at least
$6$.  Any other alternative is preferred to the Condorcet winner by at most
one voter and to the third alternative by at most four, so its score is at
most $5$.
\end{proof}

\section{Applications of the tops-only characterization}\label{sec:applications}

Having established the general coalition-size bounds and the
exceptional possibility result, we now return to the tops-only
characterization in
Theorem~\ref{thm:characterization} to determine the protection
against smaller coalitions offered by two familiar families:
generalized median voter schemes and voting by committees.
These rules are strategy-proof on their respective restricted
domains (see \citealp{BARBERA2011731}, for a survey).
Here we evaluate them on the universal domain and ask how large
a coalition must be to manipulate them obviously, and how a rule
should be chosen to protect against coalitions up to a given size.
Although only their dictatorial members are NOMG by
Theorem~\ref{thm:gsp}, the characterization allows us to determine
exactly which non-dictatorial members are protected against
coalitions of a given size. Option sets are always computed on
the universal domain $\linalt$; the restricted domains only
motivate the classes.

\subsection{Median and generalized median voter schemes}\label{subsec:gmvs}

Let $\alt=\{a,a+1,\ldots,b\}$ be a set of consecutive integers, so that
$b=a+m-1$, and let $m\geq3$.  For $x,y\in\alt$ write
$[x,y]=\{z\in\alt:x\leq z\leq y\}$.  Given $n-1$ \emph{fixed ballots}
$\alpha_1\leq\cdots\leq\alpha_{n-1}$ in $\alt$, the \emph{median voter
scheme} $f^\alpha$ selects the median of the fixed ballots and the $n$
reported tops,
\[
 f^\alpha(\bt)=\operatorname{med}\{t_1,\ldots,t_n,\alpha_1,\ldots,\alpha_{n-1}\},
\]
and we set $\alpha_0=a$ and $\alpha_n=b$.  For odd $n$, placing half of the
fixed ballots at each extreme alternative gives the median of the voters'
tops.

The option sets of a median voter scheme are obtained by counting.  Fix a
proper coalition $C$ with $c$ members and a report $\bt_C$, pool the $c$
reported tops with the $n-1$ fixed ballots, and order this list as
$z_1\leq\cdots\leq z_{n+c-1}$.  Then
\begin{equation}\label{eq:median-interval}
 O(\bt_C)=[z_c,z_n].
\end{equation}
Indeed, the outcome is the $n$th smallest of $2n-1$ numbers; it equals $z_c$
if all outsiders report $a$ and $z_n$ if all outsiders report $b$, and it is
nondecreasing in the outsiders' reports.  If all outsiders report an
alternative $x\in[z_c,z_n]$, then at least $n$ of the $2n-1$ numbers are at
most $x$ and at least $n$ of them are at least $x$, so the outcome is $x$.
In particular, if $r$ members report $b$ and the other $s=c-r$ members report
$a$, then
\begin{equation}\label{eq:median-split}
 O(\bt_C)=[\alpha_r,\alpha_{n-s}],
\end{equation}
because the pooled list consists of $s$ copies of $a$, the fixed ballots, and
$r$ copies of $b$.

\begin{proposition}\label{prop:median}
Let $m\geq3$ and let $f^\alpha$ be a median voter scheme.  Let $L$ be the
number of fixed ballots that are at most $a+1$, let $H$ be the number of
fixed ballots that are at least $b-1$, and let $E$ be the number of fixed
ballots that belong to $\{a,b\}$.  Then
\[
 K(f^\alpha)=\min\{L,H,E+1\}.
\]
Equivalently, for $k\in\{1,\ldots,n-1\}$, $f^\alpha$ is $k$-NOMG if and only
if $\alpha_k\leq a+1$, $\alpha_{n-k}\geq b-1$, and at least $k-1$ fixed
ballots belong to $\{a,b\}$.  Moreover, for $k\geq1$ there is a median voter
scheme with $K(f^\alpha)=k$ if and only if $k<n/2$ when $m\geq4$, and if and
only if $k<2n/3$ when $m=3$.
\end{proposition}

\begin{proof}
A median voter scheme is anonymous and onto, hence not dictatorial, so
$K(f^\alpha)\leq n-1$ by Theorem~\ref{thm:gsp}.  Let $q_a$ and $q_b$ be the
numbers of fixed ballots equal to $a$ and to $b$, so that $E=q_a+q_b$, and let
$C$ be a coalition with $c\leq n-1$ members.  By
Theorem~\ref{thm:characterization}, $C$ has no obvious group manipulation if
and only if it satisfies the conditional veto and every report of $C$ leaves
open one of its tops.

\emph{Step 1: if $c\leq\min\{L,H\}$, then $C$ satisfies the conditional
veto.}  By \eqref{eq:median-interval}, the lower endpoint of an option set is
largest at the report $b^C$, where it equals $\alpha_c\leq a+1$ by
\eqref{eq:median-split}, and the upper endpoint is smallest at the report
$a^C$, where it equals $\alpha_{n-c}\geq b-1$.  Hence every option set of $C$
contains $[a+1,b-1]$, and only $a$ and $b$ can be vetoed.  By
\eqref{eq:median-interval}, $a$ belongs to an option set if and only if the
pooled list contains at least $c$ copies of $a$.  Thus, if $C$ can veto $a$,
then $q_a<c$, and every report in which no member reports $a$ vetoes $a$.
The argument for $b$ is the same, counting from the other end of the list.

\emph{Step 2: if $c=\min\{L,H\}+1$, then $C$ violates the conditional veto.}
Reversing the order of the alternatives if necessary, let $L=c-1$, so that
$\alpha_c>a+1$ and $H\geq c-1$.  The report $b^C$ vetoes $a+1$, because the
lower endpoint of its option set is $\alpha_c$.  If $c=1$, the report $a^C$
has option set $[a,\alpha_{n-1}]$, which contains $a+1$ because
$\alpha_{n-1}\geq\alpha_1>a+1$.  If $c\geq2$, let one member report $b$ and
the other $c-1$ members report $a$.  By \eqref{eq:median-split}, the option
set is $[\alpha_1,\alpha_{n-c+1}]$, which contains $a+1$: $\alpha_1\leq a+1$
because $L\geq1$, and $\alpha_{n-c+1}\geq b-1\geq a+1$ because $H\geq c-1$.
In both cases a report in which no member reports $a+1$ leaves $a+1$ open.

\emph{Step 3: some report of a coalition with $c$ members leaves open none
of its tops only if $c\geq E+2$, and such a report exists if $c=E+2$.}
Suppose that $O(\bt_C)=[z_c,z_n]$ contains no top of $\bt_C$, and let $s$
members have a top below $z_c$ and $r=c-s$ members a top above $z_n$.  Both
numbers are positive: at least $c$ entries of the pooled list are at most the
largest reported top, so $z_c$ does not exceed it, and at least $c$ entries
are at least the smallest reported top, so $z_n$ is not below it.  Fewer than
$c$ entries of the pooled list are strictly below $z_c$; since $z_c>a$, these
entries include the $s$ low tops and the $q_a$ fixed ballots equal to $a$, so
$s+q_a<c$, that is, $r>q_a$.  Symmetrically, fewer than $c$ entries are
strictly above $z_n$, so $r+q_b<c$, that is, $s>q_b$.  Hence
$c=r+s\geq E+2$.  Conversely, if $c=E+2$, let $q_a+1$ members report $b$ and
$q_b+1$ members report $a$.  By \eqref{eq:median-split}, the option set is
$[\alpha_{q_a+1},\alpha_{n-q_b-1}]$, which contains neither $a$ nor $b$.

By Steps~1--3, for $k\leq n-1$ the rule is $k$-NOMG if and only if
$k\leq\min\{L,H\}$ and $k\leq E+1$.  Since $L\leq n-1$, this gives
$K(f^\alpha)=\min\{L,H,E+1\}$.  The three inequalities in the statement
express $L\geq k$, $H\geq k$ and $E\geq k-1$.

It remains to determine the attainable thresholds.  If $m\geq4$, no fixed
ballot is counted by both $L$ and $H$, so $K(f^\alpha)=k$ requires
$2k\leq L+H\leq n-1$.  Conversely, if $2k\leq n-1$, place $k$ fixed ballots
at $a$ and the other $n-1-k$ at $b$; then $L=k\leq H$ and $E+1=n$.  If $m=3$,
the fixed ballots equal to $a+1=b-1$ are counted by both $L$ and $H$, so
$L+H=2(n-1)-E$, and $L,H\geq k$ together with $E\geq k-1$ imply
$3k\leq2n-1$.  Conversely, the previous construction works if $2k\leq n-1$;
if $n/2\leq k<2n/3$, place $n-1-k$ fixed ballots at each of $a$ and $b$ and
the remaining $2k-n+1$ at $a+1$.  Then $L=H=k$ and $E=2(n-1-k)\geq k-1$.
\end{proof}

For example, if all fixed ballots are extreme alternatives, then
$K(f^\alpha)=\min\{q_a,q_b\}$, where $q_a$ and $q_b$ are the numbers of fixed
ballots at $a$ and at $b$.  Balancing the two numbers gives the largest
protection within this class: for odd $n$, the median of the tops has
$K(f^\alpha)=(n-1)/2$, exactly as plurality
(Proposition~\ref{prop:plurality}).  Placing every fixed ballot at $a$ gives
the rule that selects the smallest reported top, for which $K(f^\alpha)=0$.
The third term matters only for coalitions with at least two members: with
three alternatives and $n\geq3$, the median voter scheme with all fixed
ballots equal to $a+1$ has $L=H=n-1$ and $E=0$, so that single voters are
protected, while two voters whose tops are the two extreme alternatives
obtain $a+1$ whatever the outsiders do, and their truthful report leaves open
neither of their tops.

Median voter schemes only count how many voters report a top below a given
alternative.  Generalized median voter schemes allow the identities of these
voters to matter.  A \emph{monotonic family of fixed ballots} is a collection
$\beta=(\beta_S)_{S\subseteq N}$ of elements of $\alt$ such that $\beta_N=a$,
$\beta_\varnothing=b$, and $\beta_Q\leq\beta_S$ whenever $S\subseteq Q$.  The
\emph{generalized median voter scheme} $f^\beta$ is the tops-only rule
\[
 f^\beta(\bt)=\min_{S\subseteq N}\,\max\Bigl\{\max_{j\in S}t_j,\ \beta_S\Bigr\},
\]
where the term for $S=\varnothing$ is $\beta_\varnothing=b$.  A preference
$P_i$ is \emph{single-peaked} if $x<y\leq t(P_i)$ or $t(P_i)\leq y<x$ implies
$y\,P_i\,x$.  On the domain of single-peaked preferences, the onto, tops-only
and strategy-proof rules are exactly the generalized median voter schemes,
and the anonymous ones among them are exactly the median voter schemes
\citep{moulin1980strategy}; see also \citet{barbera1993generalized}.  For
$\bt\in\alt^N$ and
$x\in\alt$ let $L_x(\bt)=\{i\in N:t_i\leq x\}$.  If a coalition $C$ is the
union of disjoint sets $A$ and $B$, one of which may be empty, $(a^A,b^B)$
denotes the report of $C$ at which the members of $A$ report $a$ and the
members of $B$ report $b$.

\begin{lemma}\label{lem:gmvs}
Let $f=f^\beta$ be a generalized median voter scheme.
\begin{enumerate}[label=(\roman*)]
\item For every $\bt\in\alt^N$ and $x\in\alt$, $f(\bt)\leq x$ if and only if
      $\beta_{L_x(\bt)}\leq x$.  Consequently, $f$ is nondecreasing in each
      top, $\min_{i\in N}t_i\leq f(\bt)\leq\max_{i\in N}t_i$, and $f$ is onto.
\item Let $C$ be a proper coalition, let $\bt_C\in\alt^C$, and write
      $\ell(\bt_C)=f(\bt_C,a^{N\setminus C})$ and
      $u(\bt_C)=f(\bt_C,b^{N\setminus C})$.  Then
      $O(\bt_C)=[\ell(\bt_C),u(\bt_C)]$, and every $x$ in this interval is the
      outcome when all outsiders report $x$.  Moreover, $\ell$ and $u$ are
      nondecreasing in $\bt_C$, $u(\bt_C)\geq\min_{i\in C}t_i$ and
      $\ell(\bt_C)\leq\max_{i\in C}t_i$.
\item Let a proper coalition $C$ be the union of disjoint sets $A$ and $B$,
      one of which may be empty.  Then
      $O(a^A,b^B)=[\beta_{N\setminus B},\beta_A]$.  In particular,
      $O(a^C)=[a,\beta_C]$ and $O(b^C)=[\beta_{N\setminus C},b]$, and $C$ can
      veto $x$ if and only if $x>\beta_C$ or $x<\beta_{N\setminus C}$.
\end{enumerate}
\end{lemma}

\begin{proof}
(i) The term for $S$ is at most $x$ if and only if $S\subseteq L_x(\bt)$ and
$\beta_S\leq x$ (for $S=\varnothing$ the first condition is vacuous).  If
$\beta_{L_x(\bt)}\leq x$, the term for $S=L_x(\bt)$ is at most $x$.
Conversely, if the term for some $S$ is at most $x$, then
$S\subseteq L_x(\bt)$ and, by monotonicity,
$\beta_{L_x(\bt)}\leq\beta_S\leq x$.  Raising a top weakly shrinks every
$L_x(\bt)$ and therefore weakly raises every $\beta_{L_x(\bt)}$, so $f$ is
nondecreasing in each top.  For $x=\max_it_i$ we have $L_x(\bt)=N$ and
$\beta_N=a\leq x$, so $f(\bt)\leq\max_it_i$; for $x<\min_it_i$ we have
$L_x(\bt)=\varnothing$ and $\beta_\varnothing=b>x$, so $f(\bt)>x$.  In
particular $f(x^N)=x$ for every $x\in\alt$.

(ii) By monotonicity every element of $O(\bt_C)$ lies in
$[\ell(\bt_C),u(\bt_C)]$.  Conversely, let $\ell(\bt_C)\leq x\leq u(\bt_C)$
and let $\bt'=(\bt_C,x^{N\setminus C})$.  Since
$L_x(\bt')=L_x(\bt_C,a^{N\setminus C})$ and
$f(\bt_C,a^{N\setminus C})\leq x$, part (i) gives $f(\bt')\leq x$.  For $y<x$
we have $L_y(\bt')=\{i\in C:t_i\leq y\}=L_y(\bt_C,b^{N\setminus C})$, and
$f(\bt_C,b^{N\setminus C})=u(\bt_C)>y$, so part (i) gives $f(\bt')>y$.  Hence
$f(\bt')=x$.  The monotonicity of $\ell$ and $u$ and the two bounds follow
from part (i), applied to $(\bt_C,a^{N\setminus C})$ and
$(\bt_C,b^{N\setminus C})$.

(iii) At a profile at which the voters in a set $D$ report $a$ and all other
voters report $b$, we have $L_y=D$ for every $y$ with $a\leq y<b$, so by
part~(i) the outcome is $\beta_D$.  Applying this with all outsiders at $a$
and with all outsiders at $b$ gives $\ell(a^A,b^B)=\beta_{N\setminus B}$ and
$u(a^A,b^B)=\beta_A$, and the formula follows from part~(ii).  Taking
$B=\varnothing$ and $A=\varnothing$ gives $O(a^C)$ and $O(b^C)$.  By
part~(ii), $u(\br_C)\geq u(a^C)=\beta_C$ and
$\ell(\br_C)\leq\ell(b^C)=\beta_{N\setminus C}$ for every $\br_C$.  Hence $x$
belongs to every option set of $C$ if $\beta_{N\setminus C}\leq x\leq\beta_C$,
while $a^C$ vetoes every $x>\beta_C$ and $b^C$ vetoes every
$x<\beta_{N\setminus C}$.
\end{proof}

A median voter scheme is the generalized median voter scheme with
$\beta_S=\alpha_{n-|S|}$, which is a monotonic family of fixed ballots
because $\alpha_0=a$ and $\alpha_n=b$.  Indeed, the median is at most $x$ if
and only if at least $n-|L_x(\bt)|$ fixed ballots are at most $x$, that is,
if and only if $\alpha_{n-|L_x(\bt)|}\leq x$, which by
Lemma~\ref{lem:gmvs}(i) is the condition for $f^\beta(\bt)\leq x$.

\begin{theorem}\label{thm:gmvs}
Let $m\geq3$, let $f^\beta$ be a non-dictatorial generalized median voter
scheme, and let $k\in\{1,\ldots,n-1\}$.  Then $f^\beta$ is $k$-NOMG if and
only if every coalition $C$ with $1\leq|C|\leq k$ satisfies
\begin{enumerate}[label=(\roman*)]
\item $\beta_{N\setminus C}\leq a+1$ and $\beta_C\geq b-1$; and
\item for every partition of $C$ into nonempty sets $A$ and $B$,
      $\beta_{N\setminus B}=a$ or $\beta_A=b$.
\end{enumerate}
Consequently, $K(f^\beta)$ is one less than the size of the smallest
nonempty coalitions that fail (i) or (ii).
\end{theorem}

By Lemma~\ref{lem:gmvs}(iii), condition (i) says that the coalition can veto
at most the two extreme alternatives, and condition (ii) says that when the
coalition splits its reports between the two extreme alternatives, at least
one of them remains in its option set.  The grand coalition fails (i), so the
last statement is meaningful.

\begin{proof}
Throughout, $C$ is a nonempty coalition with at most $k\leq n-1$ members, so
that $C$ is proper and Lemma~\ref{lem:gmvs} applies.  By
Theorem~\ref{thm:characterization}, $f^\beta$ is $k$-NOMG if and only if
every such coalition satisfies the conditional veto and every report of such
a coalition leaves open one of its tops.  By Lemma~\ref{lem:gmvs}(iii), condition (i)
holds for $C$ if and only if $C$ cannot veto any alternative in
$[a+1,b-1]$.

\emph{Necessity of (i).}  Let $f^\beta$ be $k$-NOMG, suppose that (i) fails
for some coalition with at most $k$ members, and let $C$ be such a coalition
that is minimal with respect to inclusion.  Let $x\in[a+1,b-1]$ be an
alternative that $C$ can veto.  Suppose first that $C$ has at least two
members, and split it into two nonempty sets $A$ and $B$.  By minimality, (i)
holds for $A$ and for $B$, so $\beta_{N\setminus B}\leq a+1$ and
$\beta_A\geq b-1$, and the option set
$O(a^A,b^B)=[\beta_{N\setminus B},\beta_A]$ contains $x$.  Since
$x\notin\{a,b\}=\tau(a^A,b^B)$, this contradicts the conditional veto.

Hence $C=\{i\}$ for some voter $i$.  Suppose first that $\beta_C\leq b-2$.
The report $a^C$ vetoes $b-1$, so by the conditional veto the report $b^C$
vetoes $b-1$ as well; since $O(b^C)=[\beta_{N\setminus C},b]$, this means
that $\beta_{N\setminus C}=b$.  Then $b^C$ vetoes $a+1$, so the report $a^C$
must veto $a+1$; since $O(a^C)=[a,\beta_C]$, this means that $\beta_C=a$.  If
instead $\beta_{N\setminus C}\geq a+2$, the same two implications, in the
reverse order, give first $\beta_C=a$ and then $\beta_{N\setminus C}=b$.  In
both cases $\beta_{\{i\}}=a$ and $\beta_{N\setminus\{i\}}=b$, so by
monotonicity $\beta_S=a$ whenever $i\in S$ and $\beta_S=b$ whenever
$i\notin S$.  By Lemma~\ref{lem:gmvs}(i), $f^\beta(\bt)\leq x$ if and only if
$t_i\leq x$, for every $\bt$ and $x$.  Hence $f^\beta(\bt)=t_i$ for every
$\bt$, and $f^\beta$ is dictatorial, a contradiction.

\emph{Necessity of (ii).}  Let $f^\beta$ be $k$-NOMG and let $C$ be
partitioned into nonempty sets $A$ and $B$.  The report $(a^A,b^B)$ must
leave open one of its tops, so $[\beta_{N\setminus B},\beta_A]$ contains $a$
or $b$; that is, $\beta_{N\setminus B}=a$ or $\beta_A=b$.

\emph{Sufficiency.}  Suppose that (i) and (ii) hold for every coalition with
at most $k$ members, and fix such a coalition $C$.

By (i), $C$ can veto at most $a$ and $b$.  If $C$ can veto $b$, then
$\beta_C<b$ by Lemma~\ref{lem:gmvs}(iii).  At any profile at which no member
of $C$ reports $b$, we have $C\subseteq L_{b-1}$ and therefore
$\beta_{L_{b-1}}\leq\beta_C\leq b-1$, so the outcome is at most $b-1$ by
Lemma~\ref{lem:gmvs}(i).  Thus every report of $C$ that avoids $b$ vetoes
$b$.  If $C$ can veto $a$, then $\beta_{N\setminus C}>a$.  At any profile at
which no member of $C$ reports $a$, we have $L_a\subseteq N\setminus C$ and
therefore $\beta_{L_a}\geq\beta_{N\setminus C}>a$, so the outcome is above
$a$.  Thus every report of $C$ that avoids $a$ vetoes $a$.  Hence $C$
satisfies the conditional veto.

Finally, suppose that some option set $O(\bt_C)=[\ell,u]$ contains no top of
$\bt_C$.  By Lemma~\ref{lem:gmvs}(ii), $u$ is at least the smallest top of
$\bt_C$ and $\ell$ is at most the largest one, so the sets
$A=\{i\in C:t_i<\ell\}$ and $B=\{i\in C:t_i>u\}$ are nonempty, and they
partition $C$.  At the profile $(\bt_C,b^{N\setminus C})$, whose outcome is
$u<b$, we have $L_u=A$, so Lemma~\ref{lem:gmvs}(i) gives
$\beta_A\leq u<b$.  At the profile $(\bt_C,a^{N\setminus C})$, whose outcome
is $\ell>a$, we have $L_{\ell-1}=N\setminus B$, so Lemma~\ref{lem:gmvs}(i)
gives $\beta_{N\setminus B}>\ell-1\geq a$.  This contradicts (ii).  Hence every
report of $C$ leaves open one of its tops.

The last statement follows because $K(f^\beta)\leq n-1$ by
Theorem~\ref{thm:gsp}.
\end{proof}

For a median voter scheme, where $\beta_S=\alpha_{n-|S|}$, conditions (i) and
(ii) for all coalitions with at most $k$ members are the three inequalities of
Proposition~\ref{prop:median}.  Allowing the fixed ballots to depend on the
identities of the voters does not increase the largest attainable threshold:
\emph{if a non-dictatorial generalized median voter scheme is $k$-NOMG, then
$k<n/2$ when $m\geq4$, and $k<2n/3$ when $m=3$.}  Indeed, let $m\geq4$ and
suppose that $n-k\leq k$.  Choose $D$ with $|D|=n-k$; both $D$ and
$N\setminus D$ are nonempty and have at most $k$ members.  Condition (i) for
$N\setminus D$ gives $\beta_D\leq a+1$, and condition (i) for $D$ gives
$\beta_D\geq b-1$, so $m=b-a+1\leq3$, a contradiction.  Let $m=3$, so that
$a+1=b-1$, and suppose that $3k\geq2n$.  Then $2(n-k)\leq k$, so there are
disjoint coalitions $A$ and $B$ with $|A|=|B|=n-k\geq1$, and $C=A\cup B$ has
at most $k$ members.  The coalitions $N\setminus A$ and $N\setminus B$ have
$k$ members each, so condition (i) gives $\beta_A\leq a+1<b$ and
$\beta_{N\setminus B}\geq b-1>a$, and (ii) fails for $C$, a contradiction.  By
Proposition~\ref{prop:median}, both bounds are attained by median voter
schemes.

\subsection{Voting by committees}\label{subsec:committees}

Let $\Omega$ be a finite set of \emph{objects} with $|\Omega|\geq2$ (one
object is the case of two alternatives of Section~\ref{sec:binary}), and let
$\alt=2^\Omega$, so that $m\geq4$ and each top $t_i$ is a set of objects.  A
\emph{committee} for $\omega\in\Omega$ is a nonempty family
$\mathcal W_\omega$ of nonempty coalitions, the \emph{winning coalitions} for
$\omega$, such that $M\in\mathcal W_\omega$ and $M\subseteq M'$ imply
$M'\in\mathcal W_\omega$.  Given
$\mathcal W=(\mathcal W_\omega)_{\omega\in\Omega}$, \emph{voting by
committees} is the tops-only rule
\[
 f^{\mathcal W}(\bt)=\bigl\{\omega\in\Omega:\{i\in N:\omega\in t_i\}\in
 \mathcal W_\omega\bigr\}.
\]
It is onto: since $N\in\mathcal W_\omega$ and
$\varnothing\notin\mathcal W_\omega$, $f^{\mathcal W}(x^N)=x$ for every $x$.
\emph{Voting by quota} is the case
$\mathcal W_\omega=\{M\subseteq N:|M|\geq q_\omega\}$ with
$1\leq q_\omega\leq n$ for every $\omega$, and we write $f^q$.  A preference
$P_i$ is \emph{separable} if, for every $y\subseteq\Omega$ and
$\omega\notin y$, $y\cup\{\omega\}\,P_i\,y$ if and only if
$\omega\in t(P_i)$.  \citet{barbera1991voting} show that, on the domain of
separable preferences,
the onto strategy-proof rules are exactly the voting by committees, and the
anonymous ones among them are exactly the voting by quota.

We say that a coalition $C$ \emph{can force $\omega$ in} if
$C\in\mathcal W_\omega$, and that it \emph{can force $\omega$ out} if
$N\setminus C\notin\mathcal W_\omega$.  By monotonicity of the committee, if
$C$ can force $\omega$ in (out), then $\omega$ belongs (does not belong) to
the outcome whenever all members of $C$ include $\omega$ in (exclude $\omega$
from) their tops, whatever the outsiders report.

\begin{theorem}\label{thm:committees}
Let $|\Omega|\geq2$, let $f^{\mathcal W}$ be a non-dictatorial voting by
committees, and let $k\in\{1,\ldots,n-1\}$.  Then $f^{\mathcal W}$ is
$k$-NOMG if and only if no coalition with at most $k$ members can force an
object in or out.  Consequently,
\[
 K(f^{\mathcal W})=
 \min_{\omega\in\Omega}\;
 \min\bigl\{|C|:\ C\in\mathcal W_\omega
            \ \text{or}\ N\setminus C\notin\mathcal W_\omega\bigr\}-1,
\]
and $K(f^{\mathcal W})<n/2$.
\end{theorem}

\begin{proof}
\emph{Sufficiency.}  Suppose that no coalition with at most $k$ members can
force an object in or out, and let $C$ be such a coalition and $\bt_C$ one of
its reports.  If all outsiders report a set of objects $y$, the outcome is
$y$: every $\omega\in y$ is supported by a coalition that contains
$N\setminus C$, which is winning for $\omega$, and every $\omega\notin y$ is
supported by a subset of $C$, which is not.  Hence every option set of $C$ is
$\alt$, and $C$ has no obvious group manipulation.

\emph{Necessity.}  Let $f^{\mathcal W}$ be $k$-NOMG, suppose that some
coalition with at most $k$ members can force an object in or out, and let $C$
be such a coalition that is minimal with respect to inclusion.

We first show that $C$ can force every object in and out.  Suppose that $C$
can impose a decision on an object $\omega$: all sets of objects with the
opposite decision on $\omega$ are then vetoed by the same report of $C$.  Let
$y$ be one of these sets, let $\eta\neq\omega$ be another object, and let $x$
be the set obtained from $y$ by reversing the decision on $\eta$ only.  Since
$y\notin\tau(x^C)$, the conditional veto requires the report $x^C$ to veto
$y$.  If all outsiders report $y$, all voters agree on every object other
than $\eta$, and the outcome agrees with them on these objects.  Since the
outcome is not $y$, the decision on $\eta$ is the one supported by the
members of $C$, against all outsiders.  Choosing $y$ with $\eta\in y$ shows
that $C$ can force $\eta$ out, and choosing $y$ with $\eta\notin y$ shows
that $C$ can force $\eta$ in.  Applying the same argument with the roles of
$\eta$ and $\omega$ exchanged shows that $C$ can force $\omega$ in and out as
well.

If $C=\{i\}$, then for every object the winning coalitions are exactly the
coalitions that contain $i$, so $f^{\mathcal W}(\bt)=t_i$ for every $\bt$,
and $f^{\mathcal W}$ is dictatorial, a contradiction.  Otherwise, split $C$
into two nonempty sets, fix two different objects, and consider the report
at which the members of one set report the first object alone and the members
of the other set report the second object alone.  If all outsiders report the
empty set, each of the two objects is supported by a nonempty proper subset
of $C$, which by minimality is not winning, and no other object is supported
by any voter.  Hence the empty set belongs to the option set of this report,
although no member reports it; but $C$ can veto the empty set, because it can
force an object in.  This contradicts the conditional veto.

\emph{The threshold.}  The formula for $K(f^{\mathcal W})$ restates the
equivalence, because $K(f^{\mathcal W})\leq n-1$ by Theorem~\ref{thm:gsp}.
For the bound, take a coalition with $\lfloor n/2\rfloor$ members and an
object $\omega$.  If the coalition is winning for $\omega$, it can force
$\omega$ in; otherwise its complement, which has $\lceil n/2\rceil$ members,
can force $\omega$ out.  Hence some coalition with at most $\lceil n/2\rceil$
members can force an object in or out, and
$K(f^{\mathcal W})\leq\lceil n/2\rceil-1<n/2$.
\end{proof}

\begin{corollary}\label{cor:quota}
Let $|\Omega|\geq2$ and let $f^q$ be a voting by quota.  Then
\[
 K(f^q)=\min_{\omega\in\Omega}\min\{q_\omega-1,\,n-q_\omega\}.
\]
Equivalently, for $k\in\{1,\ldots,n-1\}$, $f^q$ is $k$-NOMG if and only if
$k+1\leq q_\omega\leq n-k$ for every $\omega\in\Omega$.
\end{corollary}

\begin{proof}
Voting by quota is anonymous and onto, hence not dictatorial.  The smallest
coalitions that can force $\omega$ in have $q_\omega$ members, and the
smallest coalitions that can force $\omega$ out have $n-q_\omega+1$ members.
Apply Theorem~\ref{thm:committees}.
\end{proof}

The object with the lowest threshold determines the protection of the whole
rule.  A unanimity quota on a single object, $q_\omega=n$, already gives
$K(f^q)=0$.  If every quota is $\lceil2n/3\rceil$, then
$K(f^q)=\lfloor n/3\rfloor$.  By Theorem~\ref{thm:committees}, no
non-dictatorial voting by committees protects coalitions with more than
$h=\lceil n/2\rceil-1$ members, and by Corollary~\ref{cor:quota} the voting
by quota that attain $h$ are precisely those with $h+1\leq q_\omega\leq n-h$
for every object.  Thus an odd electorate has a unique optimal quota for each
object, the simple majority quota $(n+1)/2$, while an even electorate admits
either of the two quotas $n/2$ and $n/2+1$, independently across objects.

\begin{remark}\label{rem:ab}
For $k=1$, the results of this section are those obtained by
\citet{ARRIBILLAGA202412} for individual manipulations.
Proposition~\ref{prop:median} gives $\alpha_1\in\{a,a+1\}$ and
$\alpha_{n-1}\in\{b-1,b\}$; Theorem~\ref{thm:gmvs} gives
$\beta_{N\setminus\{i\}}\in\{a,a+1\}$ and $\beta_{\{i\}}\in\{b-1,b\}$ for
every $i$; Corollary~\ref{cor:quota} gives $2\leq q_\omega\leq n-1$; and
Theorem~\ref{thm:committees} says that, for every object, no voter belongs to
all winning coalitions and no voter is a winning coalition by herself.  The
coalitional conditions are more demanding in two ways.  Conditions on single
voters become conditions on all coalitions with at most $k$ members, and, for
generalized median voter schemes, condition (ii) of Theorem~\ref{thm:gmvs} appears, which has no
counterpart for individuals.
\end{remark}

\section{Concluding remarks}\label{sec:conclusion}

The informational characterization in
Proposition~\ref{prop:indistinguishability} gives a precise basis
for studying obvious group manipulations: they are exactly the
deviations identifiable as group manipulations from option sets
alone. The results that follow show both the restrictions imposed
by this requirement and the scope for protection against it,
depending on the rules, the domain, and the size of the coalition.

For tops-only rules with at least three alternatives, immunity to obvious group manipulations has
the same bite as group strategy-proofness (Theorem~\ref{thm:gsp}).
\citet{doi2026obvious} find a similar pattern in housing markets, where the
same notion, together with individual rationality and pair efficiency,
singles out the top trading cycles mechanism, which is group strategy-proof.
The coincidence is not general: there are NOMG rules that are not
strategy-proof, namely the parity rules of Example~\ref{ex:parity} and, by
the Gibbard--Satterthwaite theorem, the scoring rules of
Theorem~\ref{thm:n4m3}.  Identifying the
classes of rules and the environments in which the two requirements coincide
is a natural question.

The coalition-size results also distinguish the two classes.
Efficient, non-dictatorial tops-only rules can protect all
coalitions of up to $n-2$ voters, whereas under the majority
criterion a minimal strict majority already suffices for an
obvious manipulation, outside the exceptional configuration.
For odd electorates the latter threshold is exact for every such
rule. For even electorates, determining which rules also protect
coalitions of exactly $n/2$ voters remains a question for further
research, both within the majority class and within its
Condorcet-consistent subclass.

The results for tops-only rules use the richness of the universal domain at two points: to choose the lower ranks of the members of a
coalition in Lemma~\ref{lem:tops-reform}, and to find a third alternative in
the proof of Theorem~\ref{thm:gsp}.  On restricted domains, such as the
domain of single-peaked preferences, where non-dictatorial group
strategy-proof rules exist, NOMG and (group)
strategy-proofness are not equivalent, as the following example shows.

\begin{example}\label{ex:single-peaked-nomg}
Let $\alt=\{a,b,c\}$, let $n\geq3$ be odd, and restrict preferences to
the domain of single-peaked preferences with respect to the axis
$a<b<c$: true preferences, reports and option sets are all taken within
this domain. Consider the tops-only rule that selects $c$ if all voters
rank $c$ first, $a$ if an odd number rank $a$ first, and $b$ otherwise.
This rule is anonymous and, on this domain, efficient. For every
nonempty proper coalition $C$,
\[
O^f(P_C)=
\begin{cases}
\{a,b,c\}, & \text{if all members of $C$ rank $c$ first},\\
\{a,b\}, & \text{otherwise}.
\end{cases}
\]
Since $a$ and $b$ remain attainable after every report, neither
can witness a worst-case obvious manipulation; $c$ cannot do so
either, since it is attainable truthfully only when all members
rank it first. A best-case obvious manipulation would require
making $c$ attainable when some member prefers $a$ or $b$ to it,
which also fails. Efficiency handles the grand coalition,
so the rule is NOMG. However, if two voters have the preference
$a\,b\,c$ and all others have $b\,a\,c$, the
outcome is $b$, and either of the first two voters can obtain
$a$ by reporting $b\,a\,c$. Thus, the rule is not
strategy-proof. On the universal domain the same rule is not
efficient: if two voters have the preference $a\,c\,b$ and all others
have $c\,a\,b$, it selects $b$, which every voter ranks below $a$.
This is why the example does not contradict Theorem~\ref{thm:gsp}.
\end{example}

Section~\ref{sec:applications} takes the
complementary view: it keeps the universal domain and asks how much
protection survives for rules that are strategy-proof on single-peaked or on
separable preferences.  With at least four alternatives, no non-dictatorial
rule in these classes is protected against coalitions with half of the
voters, while the best of them are protected against all smaller coalitions.

Beyond tops-only rules, Subsection~\ref{subsec:majority} and Section~\ref{sec:possibility}
establish, respectively,
the incompatibility of NOMG with the majority criterion for
$n,m\geq3$ and $(n,m)\neq(4,3)$, and the existence of anonymous
NOMG scoring rules in the exceptional configuration. Borda's rule
also shows that NOMG and Condorcet consistency can coexist there.

A characterization of the NOMG scoring rules for arbitrary numbers of voters
and alternatives is left for future work.

\bibliographystyle{apalike}
\bibliography{ref}

\clearpage
\appendix
\section{Proof of Theorem~\ref{thm:n4m3}}\label{app:n4m3}

Throughout, $n=4$, $m=3$, $s=(1,t,0)$, and $\Sc(x)$ denotes the total score
of $x$.  For a coalition $C$ with $c=|C|$ members and a report $P_C$,
$\sigma(x)$ denotes the score that $x$ receives from the members of $C$, and
$r=4-c$ is the number of outsiders.  We use three observations.
\begin{enumerate}[label=(\alph*)]
\item \emph{Exclusion.}  If $\sigma(y)-\sigma(x)>r$ for some $y$, then
      $x\notin O(P_C)$, since each outsider gives $x$ at most $1$ point more
      than $y$.
\item \emph{Large coalitions.}  If $c\cdot\min\{t,1-t\}>r$, then $C$ has no
      group manipulation.  Indeed, if the outcome changes from $x=f(P)$ to
      $y$ with $y\,P_i\,x$ for every $i\in C$, then at $P$ each member gives
      $y$ at least $\min\{t,1-t\}$ more than $x$, so the members give $y$
      more than $r$ points more than $x$, while the outsiders give $x$ at
      most $r$ more than $y$; hence $\Sc(y)>\Sc(x)$ at $P$, contradicting
      $f(P)=x$.
\item \emph{Raising the target.}  If $x$ wins at some outsider profile, it
      still wins after every outsider moves $x$ to the top of her ranking:
      the score of $x$ weakly increases, every other score weakly decreases,
      and the ties won by $x$ are still won.  So, when checking whether
      $x\in O(P_C)$, we may assume that all outsiders rank $x$ first.
\end{enumerate}

\begin{lemma}\label{lem:n4-c1}
If $t<\tfrac34$, every option set of a single voter is $\alt$.
\end{lemma}

\begin{proof}
Let the voter report $p_1\,p_2\,p_3$, in this order.  If all three outsiders
report $p_1\,p_2\,p_3$, the scores of $(p_1,p_2,p_3)$ are $(4,4t,0)$; if all
report $p_2\,p_3\,p_1$, they are $(1,3+t,3t)$; if all report $p_3\,p_2\,p_1$,
they are $(1,4t,3)$.  The respective alternatives win strictly (the last one
because $t<\tfrac34$).
\end{proof}

\begin{lemma}\label{lem:n4-c2}
Let $c=2$ and $\tfrac12\leq t<\tfrac23$.
\begin{enumerate}[label=(\roman*)]
\item If the two members rank different alternatives last, then
      $O(P_C)=\alt$.
\item If both members rank $u$ last, then $O(P_C)=\alt\setminus\{u\}$ when
      $t>\tfrac12$ or $u$ is not $\pri$-maximal, and $O(P_C)=\alt$ otherwise.
\end{enumerate}
In particular, every option set of a two-voter coalition contains both
members' tops and at least two alternatives, and an alternative can be
missing only if both members rank it last and, when $t=\tfrac12$, it is not
$\pri$-maximal.
\end{lemma}

\begin{proof}
Let $x$ be an alternative that at least one member does not rank last, so
that $\sigma(x)\geq t$, and let both outsiders rank $x$ first, which gives
$\Sc(x)\geq2+t$.  If neither of the other two alternatives receives $2$
points from the members, each of them receives at most $1+t$; if the
outsiders rank one of them second each, both end with at most $1+2t<2+t$.
If one of them, say $p$, receives $2$ points from the members, both members
rank $p$ first; the remaining alternative $q$ then receives at most $t$
points from the members, because at least one member ranks $x$ second.  If
both outsiders rank $q$ second, the scores of $p$ and $q$ are at most $2$ and
$3t$, both below $2+t$.  Hence $x\in O(P_C)$.  This proves (i) and shows
that, in (ii), the two alternatives other than $u$ belong to $O(P_C)$.

It remains to consider $u$ in case (ii).  Since both members rank $u$ last,
$\Sc(u)\leq2$, while the three scores add up to $4(1+t)$.  If $t>\tfrac12$,
the scores of the other two alternatives add up to more than $4$, so one of
them exceeds $2$ and $u$ is not selected.  If $t=\tfrac12$, the three scores
add up to $6$, so $u$ can be selected only if all three scores equal $2$,
and then only if $u$ is $\pri$-maximal.  Such a tie can be obtained.  Let $p$
and $q$ be the other two alternatives.  If both members rank $p$ first, let
both outsiders report $u\,q\,p$; if the members rank $p$ and $q$ first,
respectively, let the outsiders report $u\,p\,q$ and $u\,q\,p$.
\end{proof}

\begin{proof}[Proof of Theorem~\ref{thm:n4m3}]
\emph{The case $t<\tfrac12$.}  Label the alternatives so that $w$ is
$\pri$-minimal and let $p,q$ be the others.  Let $C=\{1,2\}$, let voter~$1$
have the preference $p\,q\,w$ and voter~$2$ the preference $q\,p\,w$, so that
$\sigma=(1+t,1+t,0)$ on $(p,q,w)$.  If the outsiders report $w\,p\,q$ and
$w\,q\,p$, the scores are $(1+2t,1+2t,2)$ and $w$ wins, because
$t<\tfrac12$; so $w\in O(P_C)$.  Let both members report $p\,q\,w$, so that
$\sigma=(2,2t,0)$.  By (c), $w$ can win only if both outsiders rank it first,
which gives $w$ the score $2$ and $(p,q)$ the scores $(2,4t)$, $(2+t,3t)$ or
$(2+2t,2t)$.  In each case $p$ has at least $2$ points, and $p$ wins a tie
with $w$ because $w$ is $\pri$-minimal.  Hence $w$ is not in the option set
of the new report, and (W) holds with $w$, which both members rank last.

\emph{The case $\tfrac23\leq t<1$.}  Label the alternatives $p_1,p_2,p_3$
with $p_2\pri p_1$, and let the members of $C=\{1,2,3\}$ have the common
preference $p_1\,p_2\,p_3$, so that $\sigma=(3,3t,0)$.  If the outsider
reports $p_2\,p_3\,p_1$, the scores are $(3,3t+1,t)$ and $p_2$ wins (by
priority when $t=\tfrac23$), so $p_2\in O(P_C)$.  Let one member report
$p_1\,p_3\,p_2$ while the others keep their ballots; then
$\sigma=(3,2t,t)$, and since $3-2t>1$ and $3-t>1$, (a) gives the option set
$\{p_1\}$.  Condition (W) holds with $p_2$.

\emph{The case $t=1$.}  Let $x$ be $\pri$-maximal.  If every voter ranks some
$y\neq x$ first and $x$ second, then $x$ and $y$ tie with score $4$ and $x$
is selected, although every voter prefers $y$.  The rule is not efficient,
hence OMG by Lemma~\ref{lem:basic}(iii).

\emph{The case $\tfrac12\leq t<\tfrac23$.}  For $c=1$, all option sets equal
$\alt$ (Lemma~\ref{lem:n4-c1}), so neither (W) nor (B) can hold.  For
$c\in\{3,4\}$, (b) with $\min\{t,1-t\}=1-t$ shows that there is no group
manipulation, because $3(1-t)>1$; by Lemma~\ref{lem:basic}(i) there is no
obvious group manipulation either.  Let $c=2$, with truthful option set $T$
and deviating option set $F$.  By Lemma~\ref{lem:n4-c2}, $T$ contains both
members' tops, so (B) cannot hold: no alternative is ranked by a member above
her own top.  Suppose that (W) holds with $x\in T$.  Then $x\notin F$, and
both remaining alternatives belong to $F$, because by Lemma~\ref{lem:n4-c2}
an option set of a two-voter coalition misses at most one alternative.  Since
every element of $F$ is ranked above $x$ by both members, both members rank
$x$ last.  As $x\in T$, Lemma~\ref{lem:n4-c2}(ii), applied to the truthful
report, gives $t=\tfrac12$ and $x$ $\pri$-maximal.  But then, by
Lemma~\ref{lem:n4-c2}, no report of the coalition has an option set that
misses $x$, contradicting $x\notin F$.  Hence there is no obvious group
manipulation.
\end{proof}

\end{document}